\documentclass[11pt]{article}

\usepackage{amsfonts}
\usepackage{appendix}

\usepackage{amsmath}
\usepackage{amssymb}
\usepackage{latexsym}
\usepackage{graphicx}
\usepackage[english]{babel}
\usepackage[font={small}]{caption}
\usepackage{mathtools}

\usepackage{amsfonts}
\usepackage{latexsym}
\usepackage{graphicx}
\usepackage[english]{babel}
\usepackage{tikz}

\begin{document}
\input epsf

\def\p{\partial}
\def\h{{1\over 2}}
\def\be{\begin{equation}}
\def\bea{\begin{eqnarray}}
\def\ee{\end{equation}}
\def\eea{\end{eqnarray}}
\def\d{\partial}
\def\la{\lambda}
\def\eps{\epsilon}
\def\b{\bigskip}
\def\m{\medskip}

\newcommand{\newsection}[1]{\section{#1} \setcounter{equation}{0}}

\def\q{\quad}

\def\h{{1\over 2}}
\def\t{\tilde}
\def\r{\rightarrow}
\def\nn{\nonumber\\}

\let\p=\partial

\newcommand\blfootnote[1]{%
  \begingroup
  \renewcommand\thefootnote{}\footnote{#1}%
  \addtocounter{footnote}{-1}%
  \endgroup
}

\begin{flushright}
\end{flushright}
\vspace{20mm}
\begin{center}
{\LARGE Lifting of states in 2-dimensional $N=4$ supersymmetric CFTs}
\\
\vspace{18mm}
{\bf   Bin Guo$^1$\blfootnote{$^{1}$guo.1281@osu.edu} and  Samir D. Mathur$^2$\blfootnote{$^{2}$mathur.16@osu.edu}
\\}
\vspace{10mm}
Department of Physics,\\ The Ohio State University,\\ Columbus,
OH 43210, USA\\ \vspace{8mm}

\vspace{8mm}
\end{center}

\vspace{4mm}

\thispagestyle{empty}
\begin{abstract}

\vspace{3mm}

We consider states of the D1-D5 CFT where only the left-moving sector is excited. As we deform away from the orbifold point, some of these states will remain BPS while others can `lift'. The lifting can be computed by a path integral containing two twist deformations; however, the relevant 4-point amplitude cannot be computed explicitly in many cases. We analyze an older proposal by Gava and Narain where the lift can be computed in terms of a finite number of 3-point functions. A direct Hamiltonian decomposition of the path integral involves an infinite number of 3-point functions, as well the first order correction to the starting state. We note that these corrections to the state account for the infinite number of 3-point functions arising from higher energy states, and one can indeed express the path-integral result in terms of a finite number of 3-point functions involving only the leading order states that are degenerate. The first order correction to the supercharge $\bar G^{(1)}$ gets replaced by a projection $\bar G^{(P)}$; this projected operator can also be used to group the states into multiplets whose members have the same lifting.

\end{abstract}
\newpage

\setcounter{page}{1}

\numberwithin{equation}{section} 

\section{Introduction}

String theory provides a very useful model for black holes: the D1D5P extremal hole in 4+1 noncompact dimensions \cite{sv, cm, dmcompare, maldastrom}. We compactify IIB string theory as
\be
M_{9,1}\r M_{4,1}\times S^1\times {\mathcal M}_4
\ee
where ${\mathcal M}_4$ is $K3$ or $T^4$. The bound state of D1 and D5 branes generates a 1+1 dimensional CFT. The P charge is obtained by adding left moving excitations in this CFT.

In the moduli space of the CFT, it is believed that there is an `orbifold point' where the theory can be expressed in terms of  free bosons and fermions, subject only to an orbifolding constraint \cite{Vafa:1995bm,Dijkgraaf:1998gf,orbifold2,Larsen:1999uk,Arutyunov:1997gt,Arutyunov:1997gi,Jevicki:1998bm,David:2002wn}. The CFT at the orbifold point is given by a 1+1 dimensional sigma model with target space $({\mathcal M})^N/S^N$, where $S^N$ is the symmetric group. The dual gravity theory lives on $AdS_3\times S^3\times {\mathcal M}_4$. A low curvature $AdS$ emerges at values of the moduli far from the orbifold point.

\subsection{Deforming the CFT}

At the orbifold point  we can take any set of left moving boson and fermion excitations in the CFT, and we will obtain a supersymmetric state with mass equalling charge. As we move away from the orbifold point, some of these states will lift to higher energies, while others will remain supersymmetric. The states that are protected against lifting can be counted using an index. This index was computed in \cite{sv} for ${\mathcal M_4}=K3$ and in \cite{mms} for ${\mathcal M}_4=T^4$. 

The index does not however tell us the detailed pattern of lifting: i.e., which particular states lift and by how much. This detailed information is useful to understand the structure of black holes, where we can now obtain explicit constructions of several classes of D1D5P microstates in the gravity dual \cite{fuzzballs_i,fuzzballs_ii,fuzzballs_iii,fuzzballs_iv,fuzzballs_v}. 

The deformation of the CFT off the orbifold point is given by adding a deformation operator $D$ to the action
\be
S\r S+\lambda \int d^2 z D(z, \bar z)
\ee
where $D$ has conformal dimensions $(h, \bar h)=(1,1)$. A choice of $D$ which is a singlet under all the symmetries at the orbifold point is
\be\label{D 1/4}
D=\frac{1}{4}\epsilon^{\dot A\dot B}\epsilon_{\alpha\beta}\epsilon_{\bar\alpha \bar\beta} G^{\alpha(0)}_{\dot A, -\h} \bar G^{\bar \alpha(0)}_{\dot B, -\h} \sigma^{\beta \bar\beta}
\ee
where $\sigma^{\beta\bar\beta}$ is a twist operator of rank $2$ in the orbifold theory.\footnote{The notation is detailed in the Appendices.} $\bar G^{(0)}$ is the right moving supercharge operator at orbifold point.

\subsection{The lifting at $O(\lambda^2)$}

 We are interested in the states which have well defined scaling dimensions, and the values of these dimensions, as we move away from the orbifold point. In particular, we are interested in right moving Ramond ground states which at the orbifold point have dimensions $(h,\frac{ c}{24})$. It turns out that while such states receive corrections at first order in $\lambda$, the dimensions get corrections only starting at $O(\lambda^2)$.  These $O(\lambda^2)$ corrections were computed for some simple states in \cite{gz,hmz}. The computation involves pulling down two copies of the deformation operator $D$ from the action, and then integrating the positions of these two $D$ operators.  Then the change in dimension has the form
 \bea\label{moti h matrix1}
\delta^{(2)} \bar h=  \lim_{T\rightarrow \infty} -\frac{\lambda^2}{4} T^{-1}e^{E^{(0)}T}\langle O^{(0)}(T/2)|\int d^2 w_{1} D(w_{1},\bar w_{1}) \int d^2 w_{2} D(w_{2},\bar w_{2}) | O^{(0)}(-T/2)\rangle
\eea
Here $| O^{(0)}(-T/2)\rangle$ is the zeroth order state at time $\tau=-{T\over 2}$ on the  cylinder. 
The question now is: can we compute this 4-point function? 
The $D$ operators have a twist component, and one way to compute the correlator is to  passing to a covering space where these twists are \cite{lm1, lm2}. Then one encounters two cases:

\b

(i) If the covering space is a sphere, the computation can be relatively straightforward. The twist operators in the $D$ give spin fields in the cover; these spins fields can be removed by spectral flow. There are additional local operators on the cover, and one computes the correlator of these operators. The correlator factorizes into a holomorphic and an anti-holomorphic part, and since we are working with right moving Ramond ground states, the anti-holomorphic part of the correlator is universal and simple to compute. 

\b

(ii)  If the covering surface is a higher genus surface, then computing the correlator is much harder. In \cite{hmz} it was shown that the covering surface is a sphere when the first $D$ joins two copies of the CFT, and the second $D$ undoes this join. But the covering space is a torus if the first $D$ splits a multiwound copy of the CFT, and the second $D$ rejoins the components. This latter case is problematic, as it is difficulty to compute the required correlators on the torus. We again get a spin field from each $D$, but we cannot remove these by spectral flow the same way as we did on the sphere. Further, the correlator involves all the conformal blocks of the theory since it is a 1-loop partition function. Thus it is not factorized into a holomorphic and an antiholomorphic part, and the antiholomorphic blocks are not simple to find.

\b

Thus the issue is: in the case where the covering space is a torus, how should we compute the lifting of states?

\subsection{The Gava-Narain proposal}
 
In \cite{gn},  Gava and Narain  proposed a method  by which one could find this $O(\lambda^2)$ correction by a computation that involves only {\it one} deformation operator $D$. In such a computation the covering space obtained by undoing the twist will always be a sphere, and the computation will therefore not have the complications arising from higher genus surfaces.  

Let us outline the Gava-Narain method. In this subsection we will assume that we are in the NS sector, as this is the sector used in their discussion.   Suppose we start at the orbifold point with an operator $O^{(0)}$ which is a chiral primary on the right but has an arbitrary dimension on the left. The right side has dimension $\bar h$ and charge $\bar j$ satisfying $h=|\bar j|$, so $O^{(0)}$ is supersymmetric under the right moving supercharges $\bar G^{\bar\alpha{(0)}}_{\dot A}$ of the CFT:
\bea
\bar G^{\bar\alpha{(0)}}_{\dot A,-\frac{1}{2}}O^{(0)}(0)=\oint_{C_{0}} \frac{d\bar z}{2\pi i} \bar G^{\bar\alpha{(0)}}_{\dot A}(\bar z) O^{(0)}(0)=0
\eea
Such an operator $O^{(0)}$ describes the states of the D1D5P system which are extremal states at the orbifold point. Note that since $\bar G^{(0)}$ is an antiholomorphic operator, the contour $C_{0}$ in the above integral can be deformed freely, upto the point where it collides with some other operator insertion. 

To describe the corrections when we move away from the orbifold point, Gava and Narain defined the following operator 
\be\label{intro delta G}
 \delta \bar G^{\bar\alpha}_{\dot A, -\h} O^{(0)}(0)=  \oint_{C_{0}} \frac{dz}{2\pi i} \bar D^{\bar\alpha}_{\dot A}(z,\bar z) O^{(0)}(0)
\ee
Here 
\be
\bar D^{\bar\alpha}_{\dot A}(z,\bar z)=\pi G^{+(0)}_{\dot A,-\frac{1}{2}}\sigma^{-\bar\alpha}(z,\bar z)
\ee
is an operator with $(h, \bar h)=(1, \h)$ (for more details, see section \ref{sec 4}). In (\ref{intro delta G}),  the contour is  to be taken around an infinitesimal circle of radius $\epsilon$ around $z=0$. This contour {\it cannot} be deformed freely since $\bar D^{\bar\alpha}_{\dot A}$ is not a holomorphic operator. 

The expression (\ref{intro delta G}) motivates Gava and Narain's proposal to compute the second order shift in the conformal dimension
$
\delta^{(2)} h =\delta^{(2)} \bar h
$.
The proposal is that
\be
\delta^{(2)} \bar h =|\lambda|^2\left |   \delta \bar G^{\bar\alpha}_{\dot A, -\h} | O^{(0)}\rangle        \right |^2
\label{vfive}
\ee
In other words we are supposed to do the following. We start with the operator $O^{(0)}$ which is right chiral at the orbifold point, with dimensions $(h,0)$. We take the OPE of $\bar D^{\bar\alpha}_{\dot A}(z,\bar z)\, O^{(0)}(0)$, and look at the operator $O_{i}^{(0)}$ which we find as coefficient of the pole ${1\over  z}$. The norm of the state $O_{i}^{(0)}|0\rangle$ then gives (after multiplication by $|\lambda|^2$) the second order lift of the dimension of $O^{(0)}$. 

\subsection{The goal of this paper}

Clearly the Gava-Narian proposal will be very useful to compute lifting, particularly for the cases where the covering space for the correlator (\ref{moti h matrix1}) is a torus. The goal of this paper is to study some aspects of the Gava-Narain operator $ \delta \bar G$ and its relation to the perturbation of the supercharge $\bar G^{(1)}$ which we would get as as an expansion in the coupling $\bar G=\bar G^{(0)}+\lambda \bar G^{(1)}+\dots$. More precisely, we discuss the following:

\b

(a)  The operation $ \delta \bar G$ arises naturally in a path integral formulation, while the operator $\bar G^{(1)}$ arises naturally in a Hamiltonian language.  We will note that $ \delta \bar G$ is not the same as $\bar G^{(1)}$; rather it is given by a projection of $\bar G^{(1)}$ onto the appropriate subspace of right-chiral operators. 

\b

(b) The above relation between $ \delta \bar G$ and $\bar G^{(1)}$ is relevant to an issue discussed in \cite{hmz}. If we are computing the lift of the dimension to order $\lambda^2$, then should we have to worry about the corrections to the state upto order $\lambda^2$? That is, if we expand the state corresponding to the primary operator $O(\lambda)$ as $|O\rangle =|O^{(0)}\rangle+\lambda |O^{(1)}\rangle+\lambda^2|O^{(2)}\rangle+\dots$, then will the corrections $|O^{(1)}\rangle, |O^{(2)}\rangle$ be involved in the lift? In computations using the path integral (\ref{moti h matrix1})  one just uses the leading order state $|O^{(0)}\rangle$ in the initial and final states, and in \cite{hmz} is was shown that for computation of the lift upto order $\lambda^2$ this is indeed correct (higher order computations will however need the change of the state in general). 

As we will see below, if we proceed in a Hamiltonian language, the lift {\it does} involve the correction $|O^{(1)}\rangle$. But in the Hamiltonian language $\bar G^{(1)}$ and $|O^{(1)}\rangle$ appear in a way that the lift can be written in terms of the Gava-Narain operation $ \delta \bar G$ applied to the leading order state $|O^{(0)}\rangle$. Understanding this relation will be one of our goals.

\b

(c) At the orbifold point the states with $\bar h=\frac{ c}{24}$ fall into short multiplets. As we perturb away from the orbifold point, some of these short multiplets will join into long multiplets and lift to $\bar h>\bar j$. All members of this long multiplet must have the same lift $\delta \bar h$.
But how do we identify which short multiplets will join to a long multiplet? If we have one member of the long multiplet, then the supercharge will move us to other members of the multiplet. But the leading order supercharge $\bar G^{(0)}$ will kill the leading order states $|O^{(0)}\rangle$. Which operator should we use to act on  $|O^{(0)}\rangle$ to get another member of the long multiplet?

We will find a set of operators in the same class as  the operator $\delta \bar G$ used by Gava-Narain, such that acting on $|O^{(0)}\rangle$ with these operators  will produce the other states of the long multiplet in which  $|O^{(0)}\rangle$ lies.  

\b

(d) We take some care to handle singularities in the computation of the path integral. As we show in the Appendices, the integral over the position of the deformation $D$ can lead to contact terms, which may be relevant in some situations.

\subsection{The plan of the paper}

The plan of the paper is as follows:
In section \ref{sec 2}, we point out some steps that need to be understood in arriving at the  Gava-Narain proposal from a Hamiltonian description. We also list the main results we find in this paper. In section \ref{sec 3}, we give a schematic outline of the steps which yield the Gava-Narain result starting from the path integral.  In sections \ref{sec 4} and \ref{sec 5}, we develop some technical tools. In section \ref{sec 6}, we carry out the detailed steps needed for the derivation.  In section \ref{sec super multi}, we show that a suitable operator can be used to group states into super-multiplets which have the same lifting. In section \ref{sec 8}, we find the grouping of terms which relates the direct Hamiltonian expression of lifting to the expression obtained by Gava-Narain, thus resolving the difficulties noted in \ref{sec 2}. Section \ref{discussion} is a summary and discussion. In the Appendices we collect some notation and helpful relations, and also note recall the subtleties in defining perturbations of chiral algebra operators.

\b

\b

Before proceeding,  we note that there are many earlier works that study conformal perturbation theory, the lifting  of the states, the acquiring of anomalous dimensions, and the issue of operator mixing,   in particular in the context of the D1-D5 CFT;  see for example \cite{Pakman:2009mi,Avery:2010er,Burrington:2012yq,Burrington:2014yia,Burrington:2017jhh,Carson:2016uwf}. Also, for more computations in  conformal perturbation theory in two and higher dimensional CFTs see, e.g.  \cite{kadanoff,Dijkgraaf:1987jt,Cardy:1987vr,Eberle:2001jq,Gaberdiel:2008fn,Berenstein:2014cia,Berenstein:2016avf}.


\section{The expression for $\delta^{(2)}\bar h$ from the operator algebra}\label{sec 2}

We will work on the cylinder, in the Ramond sector. This is useful when studying the D1D5 system in the context of black holes; of course a spectral flow operation reproduces all computations in the NS sector. 

In the NS sector a state which is right chiral has dimensions $(h, \bar h)$ with $\bar h=\bar j$. In the Ramond sector  the corresponding state has conformal dimensions $(h, \bar h)=(h, {c\over 24})$, so the right moving sector is in the Ramond ground state.  Let us further assume for the moment that this unperturbed state $|O^{(0)}\rangle$ is nondegenerate, so that we do not have to deal with the issue of operator mixing.

We have the algebra relation 
\be
\{ \bar G^+_{-, 0}, \bar G^-_{+, 0}\}=\bar L_0-{c\over 24}
\label{veightq}
\ee
Let us write
\bea
|O\rangle &=& |O^{(0)}\rangle+ \lambda|O^{(1)}\rangle+\lambda^2 |O^{(2)}\rangle+\dots \nn
\bar G^+_{-,0}&=&\bar G^{+(0)}_{-,0}+\lambda \bar G^{+(1)}_{-,0} +\lambda^2 \bar G^{+(2)}_{-,0}+\dots\nn
\bar G^-_{+,0}&=&\bar G^{-(0)}_{+,0}+\lambda \bar G^{-(1)}_{+,0} +\lambda^{2} \bar G^{-(2)}_{+,0}+\dots
\label{zzel}
\eea
where $\lambda$ is a real parameter. We have used that $ ( \bar G^+_{-,0} ) ^\dagger = \bar G^-_{+,0}$; this relations holds for each order, so  $ ( \bar G^{+(i)}_{-,0} ) ^\dagger = \bar G^{-(i)}_{+,0}$. Since $|O^{(0)}\rangle$ is a right moving Ramond ground state, we have
\be
\bar G^{+(0)}_{-,0}|O^{(0)}\rangle=0, ~~
\bar G^{-(0)}_{+,0}|O^{(0)}\rangle=0,
\ee
Thus the states  $\bar G^{+}_{-,0}|O\rangle$ and $\bar G^{-}_{+,0}|O\rangle$ start from order $O(\lambda^1)$,
\bea
\bar G^{+}_{-,0}|O \rangle &=&\lambda \left(\bar G^{+(1)}_{-,0}|O^{(0)}\rangle+\bar G^{+(0)}_{-,0}|O^{(1)}\rangle\right)+\dots \nn
\bar G^{-}_{+,0}|O \rangle &=&\lambda \left(\bar G^{-(1)}_{+,0}|O^{(0)}\rangle+\bar G^{-(0)}_{+,0}|O^{(1)}\rangle\right)+\dots
\eea
Thus upto order $O(\lambda^2)$ we have
\bea
\delta \bar h &=&\langle O| \{ \bar G^+_{-, 0}, \bar G^-_{+, 0}\}|O\rangle
=|\bar G^-_{+, 0}|O\rangle|^2+|\bar G^+_{-, 0}|O\rangle|^2\nn
&=&\lambda^2\left(\Big|\bar G^{-(1)}_{+,0}|O^{(0)}\rangle+\bar G^{-(0)}_{+,0}|O^{(1)}\rangle\Big|^2
+\Big|\bar G^{+(1)}_{-,0}|O^{(0)}\rangle+\bar G^{+(0)}_{-,0}|O^{(1)}\rangle\Big|^2\right)+\dots
\eea
Define
\bea\label{Xi}
|\Xi_1\rangle= \bar G^{-(1)}_{+,0}|O^{(0)}\rangle+\bar G^{-(0)}_{+,0}|O^{(1)}\rangle\nn
|\Xi_2\rangle= \bar G^{+(1)}_{-,0}|O^{(0)}\rangle+\bar G^{+(0)}_{-,0}|O^{(1)}\rangle
\eea
Thus we get
\be
\delta^{(2)} \bar h=\lambda^2 \left ( \left | |\Xi_1\rangle \right |^2 + \left | |\Xi_2\rangle \right |^2  \right )
\label{vten}
\ee
 We see that the states $|\Xi_1\rangle, |\Xi_2\rangle$ do involve the first order corrections $|O^{(1)}\rangle$ to the state $|O\rangle$. Further, they will involve states of many different dimensions, not just states matching the dimension of $|O^{(0)}\rangle$. Thus the expression (\ref{vten}) looks very different from the simple expression (\ref{vfive})\footnote{We will write (\ref{vfive}) more explicitly below in (\ref{zzten}) in the Ramond sector.}  which required us to project onto states with the same dimension as $|O^{(0)}\rangle$. 

\subsection{The Gava-Narain steps}

Let us now compare the above derivation of the lifting $\delta^{(2)} \bar h$ to the steps followed in \cite{gn}. We split their argument into three steps; the relation between these steps will be central to the goals of the present paper.

\b

(1) In \cite{gn} we start with the algebra
\be
\{ \bar G^+_{-, 0}, \bar G^-_{+, 0}\}=\bar L_0-{ c\over 24}
\label{veight}
\ee
where we have spectral flowed their operators to the Ramond sector. 

\b

(2)  We can take the expectation value of each side with the state $|O^{(0)}\rangle$ whose lifting we wish to compute.  The LHS has two terms arising from the anticommutator. We can insert a complete set of states between the two supercharges:
\bea\label{zzone}
A&\equiv& \langle O^{(0)} | \{\bar G^+_{-, 0},   \bar G^-_{+, 0}\}|O^{(0)}\rangle\nn
&=&\sum_i \, \langle O^{(0)} | \bar G^+_{-, 0}|O^{(0)}_i\rangle\langle O^{(0)}_i|  \bar G^-_{+, 0}|O^{(0)}\rangle+ \sum_i \, \langle O^{(0)} | \bar G^-_{+, 0}|O^{(0)}_i\rangle\langle O^{(0)}_i|  \bar G^+_{-, 0}|O^{(0)}\rangle
\eea

\b

(3) The sum over the index $i$  in (\ref{zzone}) is in general an infinite sum, involving states $|O_i^{(0)}\rangle$ with an arbitrarily high dimension. But in \cite{gn} this is replaced by a finite sum as follows. We let the ket $|O^{(0)}\rangle$ be placed at the origin $z=0$ of the plane, and the bra $\langle O^{(0)}|$ at infinity. We will be working on the cylinder with coordinate $w$ defined by $z=e^w$; on this cylinder the ket $|O^{(0)}\rangle$ be placed at  $w=-\infty =0$ of the plane, and the bra $\langle O^{(0)}|$ at $w=\infty$. The supercharge acting on $|O^{(0)}\rangle$ is applied at $w=-\infty$, just above the location where the state $|O^{(0)}\rangle$ is placed. Similarly, the acting on $\langle O^{(0)}|$ is applied at $w=\infty$, just below the location where the state $\langle O^{(0)}|$ is placed. Thus instead of (\ref{zzone}) we consider the amplitude
\bea
A'&\equiv& 
\sum_i \, \langle O^{(0)}(\infty) | \bar G^+_{-, 0}|O^{(0)}_i\rangle\langle O^{(0)}_i|  \bar G^-_{+, 0}|O^{(0)}(-\infty)\rangle\nn
&&+  \sum_i \, \langle O(\infty) | \bar G^-_{+, 0}|O^{(0)}_i\rangle\langle O^{(0)}_i|  \bar G^+_{-, 0}|O^{(0)}(-\infty)\rangle
\label{zzonep}
\eea
Consider the state at $w=-\infty$, arising from, say, the first term of the anticommutator
\be
 \bar G^-_{+, 0}|O^{(0)}\rangle= \left ( \bar G^{-(0)}_{+, 0}+\lambda \bar G^{-(1)}_{+, 0}+\dots \right ) |O^{(0)}\rangle= \lambda \bar G^{-(1)}_{+, 0} |O^{(0)}\rangle+\dots
 \ee
 where we have noted that our ket is killed by the leading order supercharge. This state has components $|O_i^{(0)}\rangle$ with arbitrarily high energies. But since we need to take an inner product with the state $\langle O^{(0)} | \bar G^+_{-, 0}$ placed at $w=\infty$, we see that only states with low dimensions will survive in the sum. Since $|O^{(0)}\rangle$ is a Ramond ground state on the right with dimensions $(h, {c\over 24})$, we find that the lowest energy states in $ \bar G^-_{+, 0}|O^{(0)}\rangle$ have dimension also equal to $(h, {c\over 24})$. These lowest energy states give a finite contribution to the amplitude, while all higher energy states give a vanishing amplitude. We therefore define a projection operator ${\mathcal P}$ which projects onto the space of states with dimension $(h, {c\over 24})$.  Define the opertaors
 \be
 \bar G^{\bar \alpha (P)}_{\dot A,0}= \mathcal P ~\bar G^{\bar \alpha (1)}_{\dot A,0}  ~ \mathcal P
 \label{zzfift}
 \ee
 Then in the amplitudes in (\ref{zzonep}) we can replace
 \be
  \bar G^-_{+, 0}|O^{(0)}(-\infty)\rangle \r  \bar G^{-(P)}_{+, 0}|O^{(0)}(-\infty)\rangle
  \ee
and so on. Then the amplitude (\ref{zzonep}) becomes
\be
A'=\Big |   \bar G^{-(P)}_{+, 0}|O^{(0)}(-\infty)\rangle \Big| ^2 + \Big|   \bar G^{+(P)}_{-, 0}|O^{(0)}(-\infty)\rangle \Big| ^2
\label{zzten}
\ee
We see that the $ \bar G^{\bar \alpha (P)}_{\dot A,0}$ are just the operators appearing in (\ref{vfive}) (after we map to the Ramond sector), and the amplitude $A'$ can thus be computed without involving any higher dimension states $|O^{(0)}_i\rangle$. The proposal of \cite{gn} is that $\delta^{(2)} \bar h=\lambda^2 A'$ gives the lifting of the state $|O^{(0)}\rangle$. 

\b


We can now see the issue that we need to address. 

\b

(i) In step (1), on the LHS, the two supercharge operators are at the same time $\tau$ on the cylinder; this is, after all, how we define a commutator or anti-commutator, and the anticommutator gives the operator $\bar L_0-{ c\over 24}$ whose change will give us the lifting $\delta^{(2)} \bar h$. But  if we insert a complete set of states in this anticommutator as in (\ref{zzone}), then we get a contribution from an infinite number of $|O^{(0)}_i\rangle$ with arbitrarily high dimensions. Thus to compute the amplitude $A$ we would need to compute an infinite number of 3-point functions.

\b

(ii) In step (2), the state we have used is the leading order state $|O^{(0)}\rangle$, rather than the full state $|O\rangle$. Since we  have not used an eigenstate of $\bar L_0$,  we cannot say that the RHS gives us $\delta^{(2)}\bar h$; there could be other terms of order $\lambda^2$ arising from $|O^{(1)}\rangle$ etc. 

\b

(iii) In step (3), we do not use the amplitude $A$, but an amplitude $A'$ where the two supercharges at at different positions; namely at $\tau=\pm\infty$. This allows the  states $|O^{(0)}_i\rangle$ with dimensions $(h, { c\over 24})$ to propagate along the cylinder, and we need to compute only a finite number of the 3-point functions involving these states. But if the supercharges are at different positions then they do not give a commutator relation like (\ref{veight}), so we cannot relate $A'$ directly to the lift of conformal dimensions. 

\b

The relation between steps (1)-(3) is what we wish to clarify.

\subsection{The relations we obtain}

(a) First we derive the  expression (\ref{zzten}) proposed in \cite{gn} starting from from a second order path integral expression of the form (\ref{moti h matrix1}). At the orbifold point the states of dimension $(h, {c\over 24})$ can form a degenerate space $|O^{(0)}_a\rangle$.
Thus we get a matrix of amplitudes $\hat E^{(2)}$;  the liftings and the corresponding zero order states are the eigenvalue and eigenstate of the  lifting matrix
\be
\hat E^{(2)}=2\delta^{(2)}\bar h=2 \lambda^2   
  \Big\{  \bar G^{+(P)\dagger}_{+,0},  \bar G^{+(P)}_{+,0} \Big\} = 2 \lambda^2   
    \Big\{  \bar G^{+(P)\dagger}_{-,0},  \bar G^{+(P)}_{-,0} \Big\} 
\ee

\b

(b) One of our main goals is to understand the relation between the expression for the lift (\ref{zzten}) obtained in \cite{gn} and the expression (\ref{vten}) which we obtained by using a direct expansion (\ref{zzel}) of the states and operators. We will compute  the first order correction of states $|O^{(1)}\rangle$ and find that
\be
 \bar G^{\bar \alpha(P)}_{\dot A,0}|O^{(0)}\rangle=
\bar G^{\bar \alpha(1)}_{\dot A,0}|O^{(0)}\rangle
+\bar G^{\bar \alpha(0)}_{\dot A,0}|O^{(1)}\rangle
\label{zzthir}
\ee
This gives the relation between the formulation of \cite{gn} and the Hamiltonian expression (\ref{vten}).

\b

(c) At the orbifold point the $|O^{(0)}_a\rangle$ lie in short multiplets. The $|O^{(0)}_a\rangle$ that get lifted will join into long multiplets. Each member of the long multiplet will have the same lifting, so knowing the structure of these multiplets will reduce the number of lifting computations that we need to perform.

But looking at the leading order states $|O^{(0)}_a\rangle$, how do we find out which states will group themselves into a multiplet? The full supercharges $\bar G^{\bar \alpha}_{\dot A}$ will, of course, move us up and down a multiplet. But the leading order part $\bar G^{\bar \alpha(0)}_{\dot A}$ kills all the $|O^{(0)}_a\rangle$, and so does not help. The next correction $\bar G^{\bar \alpha(1)}_{\dot A}$ will not vanish on the $|O^{(0)}_a\rangle$, but will in general generate states of arbitrarily high energies; i.e., states that do not lie in the subspace $|O^{(0)}_a\rangle$.

We check that the projected operators $\bar G^{\bar \alpha(P)}_{\dot A}$ have the correct properties to generate multiplets of states that have the same lifting. We consider the  two raising operators
\be
\bar G^{+(P)}_{-,0},~~~~~\bar G^{+(P)}_{+,0}
\ee
The $\bar G^{P}$ and $\hat E^{(2)}$ have the following algebraic structure
\be
\{\bar G^{+(P)}_{\dot A,0},\bar G^{+(P)}_{\dot B,0}\}=0
\ee
and
\be
[\hat E^{(2)}, \bar G^{+(P)}_{\dot A,0}]=0
\ee
The first relation ensures that there are four short multiplets in a long multiplet.\footnote{In general we can have a situation where the short multiplets are defined by states which are annihilated by only two of the four supercharges $\bar G^\alpha_{\dot A}$. In the orbifold CFT of the D1D5 system the bosons can be taken to have a vanishing momentum, so the modes $\alpha_0$ annihilate the vacuum. In this case all four supercharges annihilate the states of the short multiplet.}
The second relation ensures that all the four multiplets have the same lifting.

\section{Outline of  the derivation}\label{sec 3}

In this section we outline our derivation of the Gava-Narain result via the path integral formalism. The details of these steps will be presented in the following sections. We note that similar derivations have been carried out in \cite{verlinde,gz}. We will however carry out the computation in detail since we wish to understand relations like (\ref{zzthir}) between the path integral and Hamiltonian formulations, and also to be sure of all the divergences that arise in the path-integral and how they are to be regularized.

We will find it convenient to work on the cylinder with coordinate $w$; the results can of course be mapped at any stage to the plane through the map $z=e^w$.

We begin with an expression derived in \cite{hmz} for the second order correction to the dimensions of operators which are right chiral at the orbifold point; i.e., they have $\bar h =\bar j$ in the NS sector.\footnote{As noted above, the computations of the present paper will be in the R sector. The lifting $\delta h$ does not change when we map from the NS sector to the R sector. But it is easier to work with modes of the supercharge in the R sector since $\bar G$ is periodic around the $\sigma$ circle on the cylinder, while in the NS sector $\bar G$ is antiperiodic and so we have to keep track of a branch cut. The reason we work on the cylinder rather than the plane is that the Hamiltonian formulation is more natural on the cylinder, and one of the goals of this paper is to study the relation between the operators in the Hamiltonian formulation and in the path integral language.}  Since we are just outlining the procedure here, we will suppress many indices in this section. We proceed as follows:

\b

{\bf The set-up:}  Suppose there are several right moving Ramond ground states $|O^{(0)}_a\rangle$ at the orbifold point which have the same dimensions $(h, {c\over 24})$. We place the state $|O^{(0)}_a\rangle$ near the bottom of the cylinder at $\tau=-{T\over 2}$ and another state $\langle O^{(0)}_b|$ near the top at $\tau={T\over 2}$. we then find a matrix of amplitudes computed in second order perturbation theory:
\be\label{moti h matrix}
\delta^{(2)} \bar h_{ab}\sim \lim_{T\rightarrow \infty} T^{-1}e^{E^{(0)}T}\langle O^{(0)}_{b}(T/2)|\int d^2 w_{1} D(w_{1},\bar w_{1}) \int d^2 w_{2} D(w_{2},\bar w_{2}) | O^{(0)}_{a}(-T/2)\rangle
\ee
The integrals $\int d^2w_i$ range from $\tau=-{T\over 2}$ to $\tau={T\over 2}$, and at the end we take $T\r\infty$. 

 Each deformation operator can be written as a contour integral of $\bar G^{(0)}$ surrounding the $\bar D$ operator. 
\be\label{deformation no indices}
D(w,\bar w)\sim \oint_{C_{w}} d\bar w' \bar G^{(0)}(\bar w') \bar D(w,\bar w)
\ee
Since $\bar G^{(0)}$ is an antiholomorphic operator, The contour $C_{w}$ can be moved freely upto the point where some other operator insertion is encountered. 

\b

\begin{figure}
  \includegraphics[width=\linewidth]{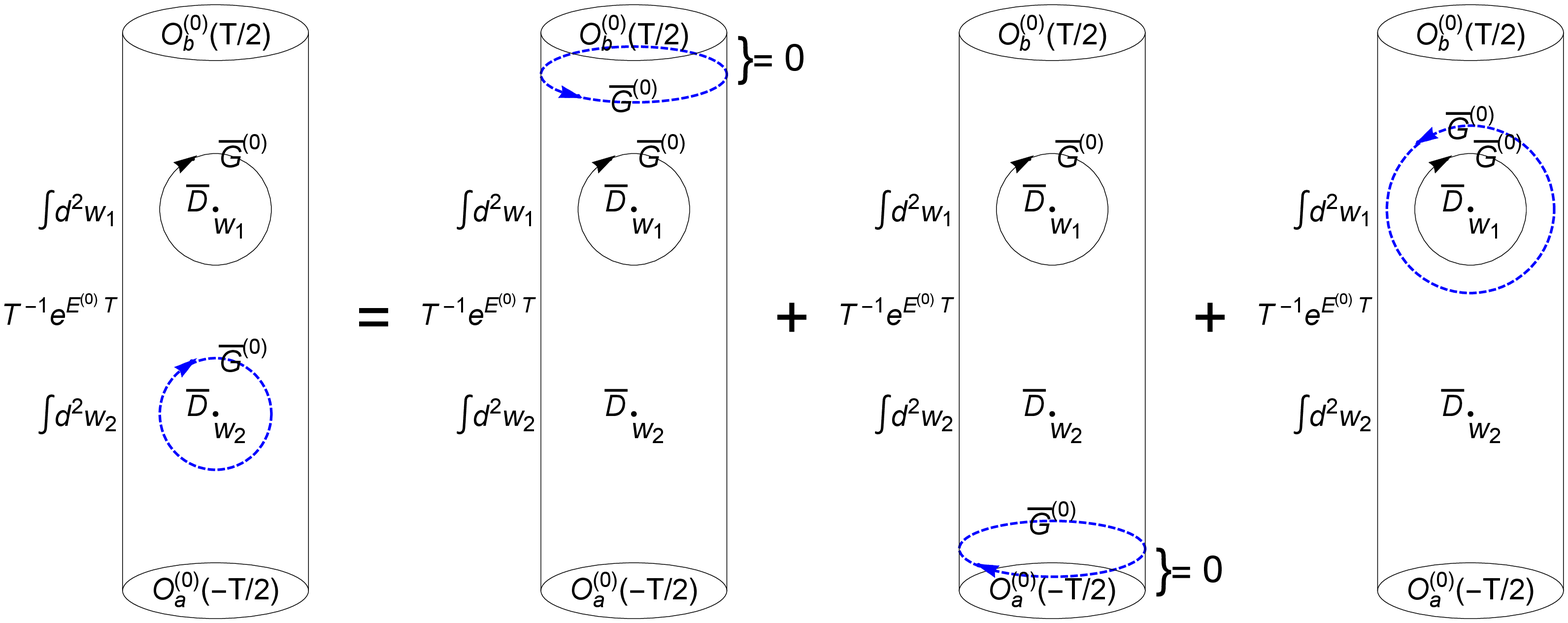}
  \caption{Step 1}
  \label{fig1}
\end{figure}

{\bf Step 1:} The manipulations are sketched in fig.\ref{fig1}.  Consider the deformation operator at $w_2$. Unwrap the $\bar G^{(0)}$ contour off this operator, and move it to the other insertions in the correlator. There are three contributions. One term gets the $\bar G^{(0)}$ contour acting on the state at the bottom $|O^{(0)}_a\rangle$. Since this state is a Ramond ground state on the right, this contribution vanishes. Similarly, the contribution from the state $\langle O^{(0)}_b|$ at the top vanishes as well. Thus the only surviving term is the third one, where the contour wraps the deformation operator at $w_1$.  

\b

\begin{figure}
  \includegraphics[width=\linewidth]{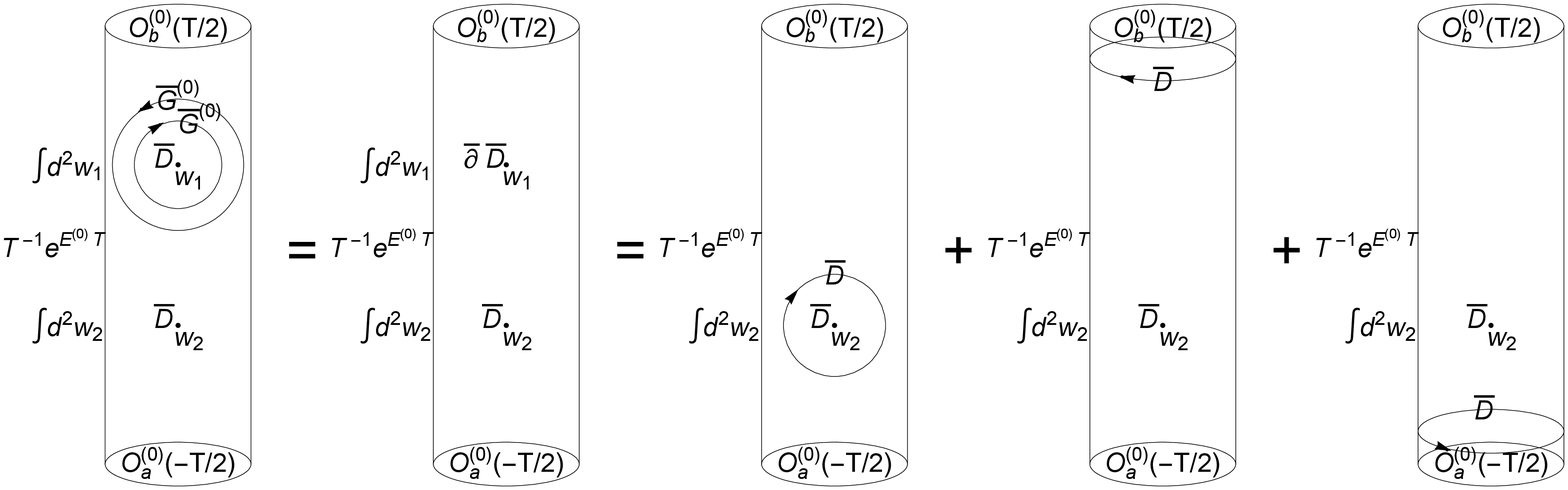}
  \caption{Step 2}
  \label{fig2}
\end{figure}

{\bf Step 2}: The manipulations are sketched in fig.\ref{fig2}. There are two $\bar G^{(0)}$ contours surrounding the insertion at $w_1$. We move the outer $\bar G^{(0)}$ contour past the inner one, getting a commutator term; the remaining contribution vanishes due to the values of the charges on the operators.  The commutator gives rise to a derivative $\p_{\bar w_{1}}$. The integral $\int d^{2}w_{1}$ can then be converted to a contour integral $\int dw_1$ of the operator $\bar D$. We get three parts to this contour. One term is around the operator insertion at $w_2$; this gives a contribution proportional to the identity operator. This contribution is divergent when the cutoff are taken to zero, but this divergent contribution has to be subtracted to keep the dimension of the identity operator to the value $0$ at second order in perturbation theory.  Thus only the other two contours are relevant. One of these contours is at $\tau=-{T\over 2}$ and the other is at $\tau={T\over 2}$. Note that the operator being integrated over these contours is not holomorphic, so we cannot freely move the contours. 

\b

{\bf Step 3:} Look at the second term we got from step 2. We still have an integral over $w_2=\tau_2+i\sigma_2$, which is $\int^{T/2}_{-T/2}d\tau_2\int_0^{2\pi} d\sigma_2$. Consider the contour at $\tau_1={T\over 2}$. Above this contour we have the state $\langle O^{(0)}_b|$ with energy $E^{(0)}=h+\bar h$. In the region $\tau_2<\tau<{T\over 2}$, the action of the contour will generate a linear combination of states with different energies. Since the contour can be written as $\int_0^{2\pi}d\sigma_1$, we see that  momentum $h-\bar h$ will remain unchanged by the action of the contour. Thus the dimensions on the left and right sides will change by the same amount.  Since the right movers are already in the ground state, the energy $E_\chi$ of the state below the contour can be equal to or larger than $E^{(0)}$, but not smaller. We will now argue that only the states with $E_\chi=E^{(0)}$ are relevant.

At the end we will be interested in the limit $T\r\infty$. The time evolution between $\tau_{2}$ and $-T/2$ gives a factor $e^{-E^{(0)}(\tau_{2}+T/2)}$. The time evolution between $T/2$ and $\tau_{2}$ gives a factor $e^{-E_\chi(T/2-\tau_{2})}$.  For $E_{\chi}>E^{(0)}$, the integral over $\tau_{2}$ gives $\sim e^{-E^{(0)}T}$, while states with $E_{\chi}=E^{(0)}$ gives  $\sim T e^{-E^{(0)}T}$. Thus only states with $E_\chi=E^{(0)}$ contribute to (\ref{moti h matrix}) in the large $T$ limit.

Thus if we are interested in the limit $T\r\infty$, it is useful to define a projection operator ${\mathcal P}$ which projects any state to the subspace with $E_\chi=E^{(0)}$. We can insert this projection operator below the contour over $w_1$, and still recover the same amplitude in the $T\r\infty$ limit. 
Doing the integral over $\tau_2$ now gives a factor $T$ which cancels the $1/T$ factor in (\ref{moti h matrix}), and we get
\be
\langle O_{b}|\int d \sigma \bar D(\sigma)\mathcal{P} \int d \sigma' \bar D(\sigma') | O_{a}\rangle
\ee
With the projection operator inserted, we can slide the contour over $w_1$ freely upto the point where it encounters the insertion at $\tau=\tau_2$, so now we can get the two contours over $\bar D$ at the same value of $\tau$; something we need to relate the amplitude to a commutation relation between operators.

\b

\begin{figure}
\centering
  \includegraphics[width=0.8\linewidth]{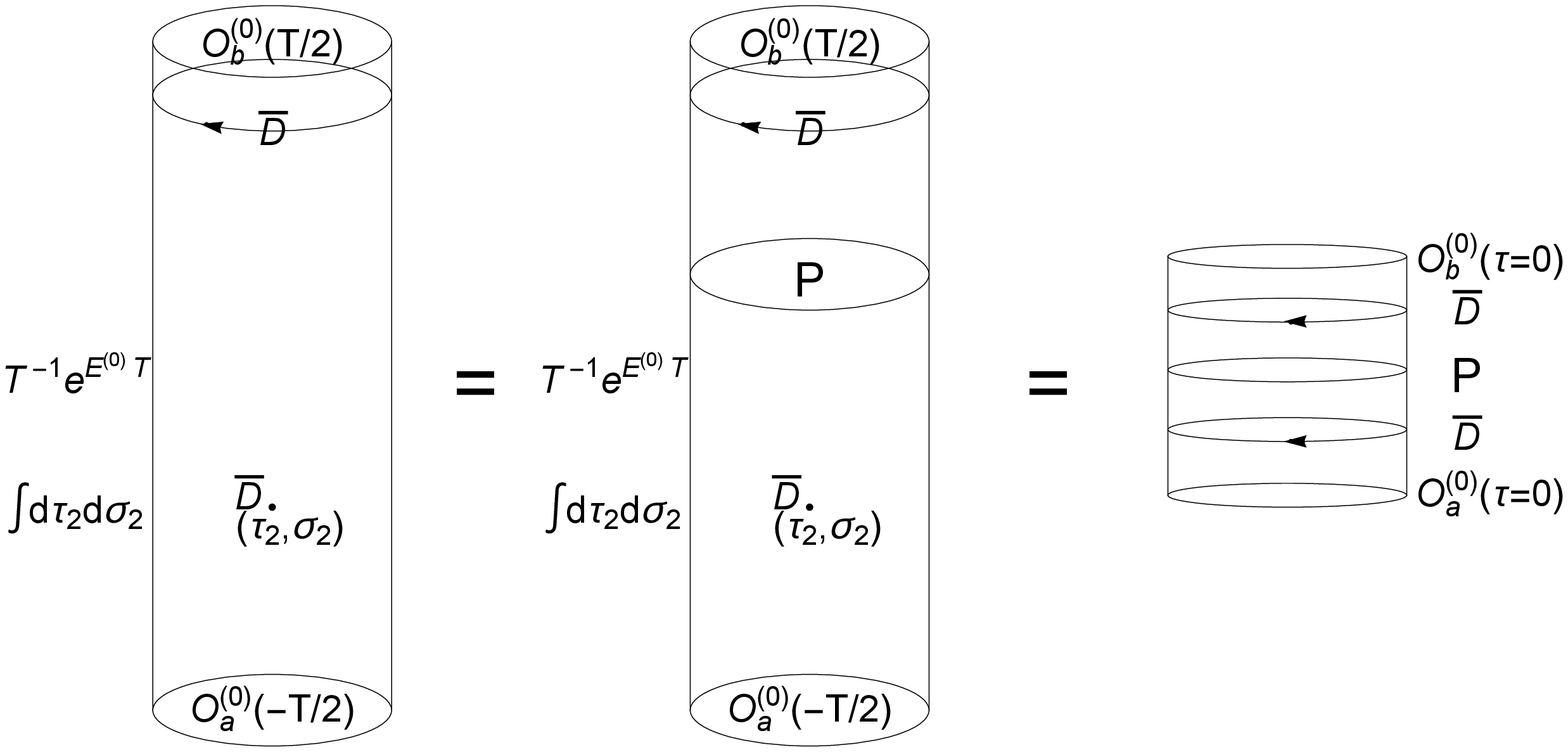}
  \caption{Step 3}
  \label{fig3}
\end{figure}

{\bf Step 4:} There is another similar term from step 2 where the contour over $w_1$ is at $\tau_1=-{T\over 2}$. We treat this term in a similar manner. It is useful to include the projection operator ${\mathcal P}$ both before and after each contour, thus defining the operators 
\be
\bar G^{(P)}\sim \mathcal{P} \int d \sigma \bar D(\sigma)\mathcal{P}
\ee
We then find that the  matrix of second order corrections to states of dimension $(h, {c\over 24})$, obtained from the path integral expression, is
\be\label{moti h matrix2}
\delta \bar h_{ab}\sim \langle O^{(0)}_{b}|\{ \bar G^{(P)},  \bar G^{(P)} \} | O^{(0)}_{a}\rangle
\ee
This expression is now just like the commutator term used by Gava-Narian. Thus we establish the correctness of their proposal, with the added fact that we must diagonalize the matrix above to find the actual change  of conformal dimensions.

\section{The operator $\bar D^{\bar \alpha}_{\dot A}$}\label{sec 4}

Let us define some technical tools which will be helpful in studying perturbations to a supersymmetric CFT. 

\subsection{Defining the operator $\bar D^{\bar \alpha}_{\dot A}$}

The CFT is defined on a 2-dimensional space. Consider a simply connected region $R$ of this space. Let $\p R$ be the boundary of $R$. 
Consider the integration of the deformation operator $D$ over this region $R$. We assume that there are no other operator insertions in $R$.
Let $\p R+\delta$ denote a curve that is a small distance $\delta$ outside the boundary of $R$. On this curve we integrate the leading order supercharge $ \bar G^{+(0)}_{\dot A}$. (We assume there are no operator insertions in the region between $\p R$ and $\p R+\delta$; the gap $\delta$ is used only as  a regulator to prevent contact between the operators $D$ and $ \bar G^{+(0)}_{\dot A}$.)
Thus we define the operator
\be\label{GD operator}
Q^{+}_{\dot A}=\oint_{\p R+\delta} \frac{d\bar w'}{2\pi i} \bar G^{+(0)}_{\dot A}(\bar w') \int_{R} d^{2}w D(w,\bar w)
\ee
We use the convention that  a contour integral for $\oint dw$ runs counterclockwise and for $\oint d\bar w$ runs clockwise.

Using the explicit expression of the deformation operator (\ref{D 1/4}),  we find
\be
Q^{+}_{\dot A}=\int_{R} d^{2}w \oint_{\p R+\delta} \frac{d\bar w'}{2\pi i} \bar G^{+(0)}_{\dot A}(\bar w')  
\epsilon^{\dot C \dot B}G^{-(0)}_{\dot C,-\frac{1}{2}}\bar G^{-(0)}_{\dot B,-\frac{1}{2}} \sigma^{++}(w,\bar w)
\ee
Using the mode expansion $\oint \frac{d\bar w'}{2\pi i} \bar G^{+(0)}_{\dot A}(\bar w')=\bar G^{+(0)}_{\dot A,-\frac{1}{2}}$, we have
\bea
Q^{+}_{\dot A}&=&\int_{R} d^{2}w \bar G^{+(0)}_{\dot A,-\frac{1}{2}}  
\epsilon^{\dot C \dot B}G^{-(0)}_{\dot C,-\frac{1}{2}}\bar G^{-(0)}_{\dot B,-\frac{1}{2}} \sigma^{++}(w,\bar w)\nn
&=&-\int_{R} d^{2}w   
\epsilon^{\dot C \dot B}G^{-(0)}_{\dot C,-\frac{1}{2}}\bar G^{+(0)}_{\dot A,-\frac{1}{2}}\bar G^{-(0)}_{\dot B,-\frac{1}{2}} \sigma^{++}(w,\bar w)
\eea
We use the relation $\bar G^{+(0)}_{\dot A,-\frac{1}{2}}\sigma^{++}=0$. (This is derived in the Appendix in eq.(\ref{A3}).)  This gives 
\be
Q^{+}_{\dot A}=-\int_{R} d^{2}w   
\epsilon^{\dot C \dot B}G^{-(0)}_{\dot C,-\frac{1}{2}}\{\bar G^{+(0)}_{\dot A,-\frac{1}{2}},\bar G^{-(0)}_{\dot B,-\frac{1}{2}}\} \sigma^{++}(w,\bar w)
\ee
Using the commutation relation (\ref{commutations_1})
\be
\{\bar G^{+(0)}_{\dot A,-\frac{1}{2}},\bar G^{-(0)}_{\dot B,-\frac{1}{2}}\}
=-\epsilon_{\dot A\dot B}\bar L^{(0)}_{-1}=-\epsilon_{\dot A\dot B}\bar\p
\ee
the operator (\ref{GD operator}) becomes
\bea
Q^{+}_{\dot A}
&=&\int_{R} d^{2}w   
\epsilon^{\dot C \dot B}G^{-(0)}_{\dot C,-\frac{1}{2}}\epsilon_{\dot A\dot B}\bar\p \sigma^{++}(w,\bar w)
=-\int_{ R} d^2 w   
\bar\p\left(G^{-(0)}_{\dot A,-\frac{1}{2}}\sigma^{++}(w,\bar w)\right)\nn
&=&-\frac{1}{2i}\oint_{\p R} dw   
G^{-(0)}_{\dot A,-\frac{1}{2}}\sigma^{++}(w,\bar w)
\eea
Here we have used $\epsilon^{\dot C \dot B}\epsilon_{\dot A\dot B}=-\delta^{\dot C}_{\dot A}$ and $\int d^2 w\bar \p=\frac{1}{2i}\int dw$.
Using $G^{-(0)}_{\dot A,-\frac{1}{2}}\sigma^{++}=-G^{+(0)}_{\dot A,-\frac{1}{2}}\sigma^{-+}$ (eq.(\ref{A4})), we have
\be\label{G1+}
Q^{+}_{\dot A}=\oint_{\p R+\delta} \frac{d\bar w'}{2\pi i} \bar G^{+(0)}_{\dot A}(\bar w') \int_{R} d^{2}w D(w,\bar w)
=\frac{1}{2i}\oint_{\p R} dw   
G^{+(0)}_{\dot A,-\frac{1}{2}}\sigma^{-+}(w,\bar w)
\ee
Note that the contour integral now runs along the exact boundary $\p R$  of $R$, not on the curve $\p R+\delta$. The integrand on this contour is not holomorphic, so this contour cannot be freely deformed.

In a similar manner we can consider the contour integral of $\bar G^{-(0)}_{\dot A}$
\bea\label{G1-}
Q^{-}_{\dot A}&=&\oint_{\p R+\delta} \frac{d\bar w'}{2\pi i} \bar G^{-(0)}_{\dot A}(\bar w') \int_{R} d^{2}w D(w,\bar w)\nn
&=&\int_{R} d^{2}w \oint_{\p R+\delta} \frac{d\bar w'}{2\pi i} \bar G^{-(0)}_{\dot A}(\bar w')  
\epsilon^{\dot C \dot B}G^{+(0)}_{\dot C,-\frac{1}{2}}\bar G^{+(0)}_{\dot B,-\frac{1}{2}} \sigma^{--}(w,\bar w)\nn
&=&-\int_{R} d^{2}w   
\epsilon^{\dot C \dot B}G^{+(0)}_{\dot C,-\frac{1}{2}}\{\bar G^{-(0)}_{\dot A,-\frac{1}{2}},\bar G^{+(0)}_{\dot B,-\frac{1}{2}}\} \sigma^{--}(w,\bar w)\nn
&=&\int_{ R} d^2 w   
\bar\p\left(G^{+(0)}_{\dot A,-\frac{1}{2}}\sigma^{--}(w,\bar w)\right)\nn
&=&\frac{1}{2i}\oint_{\p R} dw   
G^{+(0)}_{\dot A,-\frac{1}{2}}\sigma^{--}(w,\bar w)
\eea

We combine (\ref{G1+}) and (\ref{G1-}) into a single operator relation
\be\label{GD relation}
Q^{\bar\alpha}_{\dot A}=\oint_{\p R+\delta} \frac{d\bar w'}{2\pi i} \bar G^{\bar\alpha(0)}_{\dot A}(\bar w') \int_{R} d^{2}w D(w,\bar w)
= \oint_{\p R} \frac{dw}{2\pi i}   \pi
G^{+(0)}_{\dot A,-\frac{1}{2}}\sigma^{-\bar\alpha}(w,\bar w)
\ee

The definition of $Q^{\bar\alpha}_{\dot A}$  motivate us to define a local operator $\bar D^{\bar\alpha}_{\dot A}$
\be\label{D bar}
\bar D^{\bar\alpha}_{\dot A}(w,\bar w)=\pi G^{+}_{\dot A,-\frac{1}{2}}\sigma^{-\bar\alpha}(w,\bar w)
\ee
Note that this operator contains both holomorphic and anti-holomorphic parts. With this definition, the operator relation (\ref{GD relation}) becomes
\be\label{G to GN}
\oint_{\p R+\delta} \frac{d\bar w}{2\pi i} \bar G^{\bar\alpha(0)}_{\dot A}(\bar w) \int_{R} D
= \oint_{\p R} \frac{dw}{2\pi i}   
\bar D^{\bar\alpha}_{\dot A}(w,\bar w)
\ee

\begin{figure}
\centering
  \includegraphics[width=0.5\linewidth]{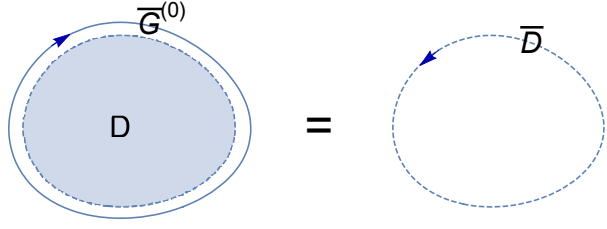}
  \caption{The picture represents eq. (\ref{G to GN}). The deformation operator is integrated over the region R (the blue region). The clockwise contour of integral $\int d\bar w$ of $\bar G^{(0)}$ is slightly outside region R, while the counterclockwise contour of integral $\int d w$ of $\bar D$ is along the boundary of R. }
  \label{fig4}
\end{figure}

Let us note the physical significance of this relation. In conformal perturbation theory, we have the integral of $D$ over the region outside the operator insertions in the correlator. Suppose we have a contour integral of the leading order supercharge $\bar G^{(0)}$ around a region $R$ over which we have the integral of $D$. Then we can replace the integral of $\bar G^{(0)}$ with the integral of $\bar D$ over the boundary of $R$, with {\it no} insertion of the deformation $D$.

Further, eq. (\ref{G to GN}) also holds for a multiply connected region $R$ which has several non-connected boundaries $\p_{i}$:
\be
\p R = \sum_{i}\p_{i}
\ee
To show this, we cut slits between the different boundaries, so that we do obtain a simply connected region where we can apply the relation (\ref{G to GN}). The integral of $\bar D$ cancels between the two edges of the slit, and we get the relation (\ref{G to GN}) again. 
In the next section, we will use this relation (\ref{G to GN}) to find the large $T$ behavior of  expressions similar to the amplitude  (\ref{moti h matrix1}).

\subsection{Deformation operator $D$ in terms of $\bar G^{\bar\alpha(0)}_{\dot A}$ and $\bar D^{\bar\beta}_{\dot B}$}

The deformation operator $D$ can be regarded as having a contour of $G^{(0)}$ and a contour of $\bar G^{(0)}$ applied to a twist operator $\sigma$. We will now rewrite $D$ as a contour of $\bar G^{(0)}$ applied to the operator $\bar D$ (\ref{D bar}).

Using (\ref{A4}), we can write the deformation operator as
\bea
D
&=&\epsilon^{\dot A\dot B}G^{+(0)}_{\dot A,-\frac{1}{2}}\bar G^{+(0)}_{\dot B,-\frac{1}{2}}\sigma^{--}
=-G^{+(0)}_{+,-\frac{1}{2}}\bar G^{+(0)}_{-,-\frac{1}{2}}\sigma^{--}+G^{+(0)}_{-,-\frac{1}{2}}\bar G^{+(0)}_{+,-\frac{1}{2}}\sigma^{--}\nn
&=&G^{+(0)}_{+,-\frac{1}{2}}\bar G^{-(0)}_{-,-\frac{1}{2}}\sigma^{-+}+G^{+(0)}_{-,-\frac{1}{2}}\bar G^{+(0)}_{+,-\frac{1}{2}}\sigma^{--}\nn
&=&-\frac{1}{\pi}\bar G^{-(0)}_{-,-\frac{1}{2}}\bar D^{+}_{+}-\frac{1}{\pi}\bar G^{+(0)}_{+,-\frac{1}{2}}\bar D^{-}_{-}
\eea
Thus we have
\be\label{deformation 1}
D(w_1,\bar w_1)=-\frac{1}{\pi}\oint_{|\bar w-\bar w_{1}|=\epsilon'} \frac{d\bar w}{2\pi i} \bar G^{-(0)}_{-}(\bar w)\bar D^{+}_{+}(w_{1},\bar w_{1})
-\frac{1}{\pi}\oint_{|\bar w-\bar w_{1}|=\epsilon'} \frac{d\bar w}{2\pi i} \bar G^{+(0)}_{+}(\bar w)\bar D^{-}_{-}(w_1,\bar w_1)\\
\ee
Similarly, we can write the deformation operator as
\bea
D
&=&\epsilon^{\dot A\dot B}G^{+(0)}_{\dot A,-\frac{1}{2}}\bar G^{+(0)}_{\dot B,-\frac{1}{2}}\sigma^{--}
=-G^{+(0)}_{+,-\frac{1}{2}}\bar G^{+(0)}_{-,-\frac{1}{2}}\sigma^{--}+G^{+(0)}_{-,-\frac{1}{2}}\bar G^{+(0)}_{+,-\frac{1}{2}}\sigma^{--}\nn
&=&-G^{+(0)}_{+,-\frac{1}{2}}\bar G^{+(0)}_{-,-\frac{1}{2}}\sigma^{--}-G^{+(0)}_{-,-\frac{1}{2}}\bar G^{-(0)}_{+,-\frac{1}{2}}\sigma^{-+}\nn
&=&\frac{1}{\pi}\bar G^{+(0)}_{-,-\frac{1}{2}}\bar D^{-}_{+}+\frac{1}{\pi}\bar G^{-(0)}_{+,-\frac{1}{2}}\bar D^{+}_{-}
\eea
This gives
\be\label{deformation 2}
D(w_1,\bar w_1)=\frac{1}{\pi}\oint_{|\bar w-\bar w_{1}|=\epsilon'} \frac{d\bar w}{2\pi i} \bar G^{+(0)}_{-}(\bar w)\bar D^{-}_{+}(w_{1},\bar w_{1})
+\frac{1}{\pi}\oint_{|\bar w-\bar w_{1}|=\epsilon'} \frac{d\bar w}{2\pi i} \bar G^{-(0)}_{+}(\bar w)\bar D^{+}_{-}(w_1,\bar w_1)\\
\ee
We will use expressions (\ref{deformation 1}) and (\ref{deformation 2}) to express a deformation operator in  terms of $\bar G^{\bar\alpha(0)}_{\dot A}$ and $\bar D^{\bar\beta}_{\dot B}$; this will be done in computing the amplitude  $\tilde X^{\bar \alpha \bar \beta}_{ba}(T)$ (\ref{X,1}) that will be defined below.

\section{Large $T$ behavior of $\tilde X^{\bar\alpha\bar\beta}_{\dot A\dot B}(T)$}\label{sec 5}

In \cite{hmz} the second order lift of right-chiral states was computed in the path integral formulation. In the Ramond sector, the unperturbed states have an arbitrary dimension $h$ on the left side and dimension $\bar h={ c\over 24}$ on the right side. Let the set of states with this dimension be $|O^{(0)}_a\rangle$. Then  one should compute the matrix elements
\be\label{X}
X_{ba}(T)=\Big\langle O^{(0)}_{b}\left(\frac{T}{2}\right)\Big|\left(\int d^{2}w_{1}D(w_{1},\bar w_{1})\right)\left(\int d^{2}w_{2}D(w_{2},\bar w_{2})\right)\Big|O^{(0)}_{a}\left(-\frac{T}{2}\right)\Big\rangle
\ee
We then compute 
\be
E^{(2)}_{ba}=\lim_{T\rightarrow \infty}-\frac{\lambda^2}{2T}e^{E^{(0)}T}X_{ba}(T)+ \frac{2\pi^2 \lambda^2}{\epsilon^2}\delta_{ba}
\ee
Here a counterterm has been added to ensure that the energy of the vacuum state is unchanged. Thus in the NS sector the energy of the vacuum $|0\rangle$ will remain zero. Correspondingly, the energy of the Ramond ground states will remain $(\frac{c}{24},\frac{ c}{24})$.  After spectral flow to the Ramond sector, we  any state that remains a Ramond ground state after perturbation  will have $h=\bar h={c\over 24}$. This counterterm is the one that arises in any second order perturbation of a CFT when the two insertions  of the deformation operator $D$ collide. We do the path integral with the prescription that the distance $|w-w'|$ between the two $D$ insertions satisfies $|w-w'|>\epsilon$. 

In our present analysis, we would like to express the first deformation operator $D(w_1, \bar w_1)$  in terms of a contour of $\bar G^{\bar \alpha(0)}_{\dot A}$ applied to $\bar D^{\bar\beta}_{\dot B}$,  by using (\ref{deformation 1}) and (\ref{deformation 2}). The deformation operator has a particular contraction between indices $\bar \alpha, \bar \beta$ and $\dot A, \dot B$. However  we will compute a more general amplitude by  these indices arbitrary for now. These general amplitudes will be needed for finding  the super multiplets structure in section \ref{sec super multi}. Thus we consider the  expression 
\bea\label{X,1}
&&(\tilde X^{\bar\alpha\bar\beta}_{\dot A\dot B})_{ba}(T)=\nn
&&\hspace{-20 pt}\int \hspace{-1 pt}d^{2}w_{1}
 \Big\langle O^{(0)}_{b}\hspace{-1 pt}\left(\frac{T}{2}\right)\hspace{-1 pt}\Big|\hspace{-1 pt}  \left(\frac{1}{\pi}\hspace{-1 pt}\oint_{|\bar w-\bar w_{1}|=\epsilon'} \hspace{-2 pt}\frac{d\bar w}{2\pi i} \bar G^{\bar\alpha(0)}_{\dot A}\hspace{-1 pt}(\bar w)\bar D^{\bar\beta}_{\dot B}(w_{1},\bar w_{1})\hspace{-1 pt}\right)\hspace{-3 pt}
\left(\int d^2 w_{2} D(w_{2},\bar w_{2})\hspace{-1 pt}\right) \hspace{-1 pt}\Big|O^{(0)}_{a}\hspace{-1 pt}\left(\hspace{-1 pt}-\frac{T}{2}\right)\hspace{-1 pt}\Big\rangle\nn
\eea
We see that by taking particular values for the indices in (\ref{X,1}), we can recover the two terms on the RHS of an expression like (\ref{deformation 1}) or (\ref{deformation 2}).

Let us now carefully consider the domain of the integrals in (\ref{X,1}):

\b

(i) The range of the $\sigma_i$ are
\be\label{region c1}
0\leq \sigma_{i}<2\pi
\ee
 except at places where a domain is cut out due to some other regularization.
 
 \b

(ii) The state $\langle O^{(0)}_{b}|$ is placed at $\tau={T\over 2}$ and the state $|O^{(0)}_a\rangle$ is at $\tau=-{T\over 2}$. We leave a small gap of width $\delta \tau=\epsilon''$ between the state insertions and the domain where the deformation operators are integrated. Thus we have
 \be\label{region c2}
 -\frac{T}{2}+\epsilon''<\tau_{i}<\frac{T}{2}-\epsilon''
 \ee
 
 \b
 
 (iii) We do not allow the two deformation operators to come within a distance $\epsilon$ of each other. Thus we have
 \be
 |w_{1}-w_{2}|>\epsilon
 \label{vel}
 \ee

\b

(iv) The integral $\oint_{|\bar w-\bar w_{1}|=\epsilon'} \frac{d\bar w}{2\pi i} \bar G^{\bar\alpha(0)}_{\dot A}(\bar w)$ is performed around a  contour surrounding $w_1$. The size of this contour is arbitrary, as long as the contour does not hit any other operator insertion We choose  this contour be a small circle with radius
\be
\epsilon'<\epsilon
\label{vtw}
\ee
With this choice there will be no collision of operators  when $w_1$ approaches $w_2$:  we have $|w_1-w_2|>\epsilon $ by (\ref{vel}), and (\ref{vtw}) then says that $\bar G^{\bar\alpha(0)}_{\dot A}(\bar w)$ will not collide with $D(w_2, \bar w_2)$.

\subsection{Simplify $\tilde X^{\bar\alpha\bar\beta}_{\dot A\dot B}(T)$}

In this subsection, we will use the relation (\ref{G to GN}) to simplify $\tilde X^{\bar\alpha\bar\beta}_{\dot A\dot B}$. Our next step is to unwrap the $\bar G^{\bar\alpha(0)}_{\dot A}(\bar w)$ from around $w_1$ and wrap it around the other insertions in the amplitude (\ref{X,1}). Since the states at the top and bottom of the cylinder are Ramond ground states of the right sector, the contour of  $\bar G^{\bar\alpha(0)}_{\dot A}(\bar w)$ around them vanishes, and we are only left with the contour around the operator insertion at $w_2$. 

While this step is straightforward, let us write it  more formally, by adding to (\ref{X,1}) the two vanishing terms 
\bea\label{2terms RH}
&&\hspace{-20 pt}-\int\hspace{-1 pt} d^{2}w_{1} \Big\langle O^{(0)}_{b}\hspace{-1 pt}\left(\frac{T}{2}\right)\hspace{-1 pt}\Big| \frac{1}{\pi} \Big(\hspace{-1 pt}\oint_{C_{\frac{T}{2}-\epsilon'''}}\hspace{-4 pt} \frac{d\bar w}{2\pi i} \bar G^{\bar\alpha(0)}_{\dot A}(\bar w)\Big) \bar D^{\bar\beta}_{\dot B}(w_{1},\bar w_{1})\Big(\int d^2 w_{2} D(w_{2},\bar w_{2})\Big) \Big|O^{(0)}_{a}\hspace{-1 pt}\left(\hspace{-1 pt}-\frac{T}{2}\right)\hspace{-1 pt}\Big\rangle\nn
&&\hspace{-20 pt}+\int\hspace{-1 pt} d^{2}w_{1} \Big\langle O^{(0)}_{b}\hspace{-1 pt}\left(\frac{T}{2}\right)\hspace{-1 pt}\Big| \frac{1}{\pi} \bar D^{\bar\beta}_{\dot B}(w_{1},\bar w_{1})\Big(\hspace{-1 pt}\int d^2 w_{2} \hspace{-1 pt}D(w_{2},\bar w_{2})\Big)\Big(\hspace{-1 pt} \oint_{C_{\hspace{-1 pt}-\frac{T}{2}+\epsilon'''}}\hspace{-4 pt} \frac{d\bar w}{2\pi i} \bar G^{\bar\alpha(0)}_{\dot A}(\bar w)\Big)\Big|O^{(0)}_{a}\hspace{-1 pt}\left(\hspace{-1 pt}-\frac{T}{2}\right)\hspace{-1 pt}\Big\rangle\nn
\eea
Here we have chosen  
\be
\epsilon'''< \epsilon''
\ee
In the first term the contour $C_{\frac{T}{2}-\epsilon'''}$ is at $\tau=\frac{T}{2}-\epsilon'''$; thus it is  between the location of the final state at $\frac{T}{2}$ and the boundary $\tau=\frac{T}{2}-\epsilon''$ of the  integration region for the deformation operators.  Similarly for the second term, the contour is between the location of the initial state and the boundary $\tau=-\frac{T}{2}+\epsilon''$ of the  integration region for the deformation operators. 

By adding (\ref{2terms RH}), which is zero, to the amplitude $\tilde X^{\bar\alpha\bar\beta}_{\dot A\dot B}(T)$ (\ref{X,1}), we have an expression where the integral of $\bar G^{\bar\alpha(0)}_{\dot A}(\bar w)$ is over a contour  which runs slightly outside the region where the deformation operator $D(w_2, \bar w_2)$ is integrated. We write this fact as
\bea\label{X to RH}
&&(\tilde X^{\bar\alpha\bar\beta}_{\dot A\dot B})_{ba}(T)=\nn
&&\hspace{-18 pt}-\int\hspace{-1 pt} d^{2}w_{1} \Big\langle O^{(0)}_{b}\hspace{-1 pt}\left(\frac{T}{2}\right)\Big| \frac{1}{\pi} \Big(\oint_{\p R+\delta} \frac{d\bar w}{2\pi i} \bar G^{\bar\alpha(0)}_{\dot A}(\bar w)\Big)\bar D^{\bar\beta}_{\dot B}(w_{1},\bar w_{1})\Big(\int_R \hspace{-1 pt} d^2 w_{2} D(w_{2},\bar w_{2})\Big) \Big|O^{(0)}_{a}\hspace{-1 pt}\left(\hspace{-1 pt}-\frac{T}{2}\right)\hspace{-1 pt}\Big\rangle\nn
\eea
where $R$ is the domain where $w_2$ is to be integrated and $\p R+\delta$ denotes a curve just outside the boundary of $R$. Recall that the region $R$ is the region of $w_{2}$ in expression (\ref{X,1}) satisfying the conditions (\ref{region c1}),(\ref{region c2}),(\ref{vel}). The clockwise boundary of $R$ is 
\be
\p R = C_{w_{1}=\epsilon}+C_{\frac{T}{2}-\epsilon''}+C_{-\frac{T}{2}+\epsilon''}
\label{vfift}
\ee
Note that  in obtaining  (\ref{X to RH}), we get a negative sign from moving the $\bar G^{\bar\alpha(0)}_{\dot A}$ operator across $\bar D^{\bar\beta}_{\dot B}$ in the second term in (\ref{2terms RH}).

The expression (\ref{X to RH}) is now of the form where we can  use the general result eq. (\ref{G to GN}) and replace the contour integral of $\bar G^{\bar\alpha(0)}_{\dot A}(\bar w)$ by a contour integral of $\bar D^{\bar\alpha}_{\dot A}$. The latter contour will run along $\p R$ given by (\ref{vfift}). We find\footnote{Notice that when $w_{1}$ is close to the upper boundary $\frac{T}{2}-\epsilon''-\epsilon<\tau_{1}<\frac{T}{2}-\epsilon''$ or close to the lower boundary $-\frac{T}{2}+\epsilon''<\tau_{1}<-\frac{T}{2}+\epsilon''+\epsilon$, the contours in (\ref{GN3terms}) are slightly different. However, as one can see by using similar method later, these regions will give a $e^{-E^{(0)}T}$ not $T e^{-E^{(0)}T}$ at large T. In the next section, we will find the large $T$ behavior for each terms in (\ref{GN3terms}).}
\bea\label{GN3terms}
&&(\tilde X^{\bar\alpha\bar\beta}_{\dot A\dot B})_{ba}(T)\nn
&=&-\frac{1}{\pi}\int d^{2}w_{1} \Big\langle O^{(0)}_{b}\left(\frac{T}{2}\right)\Big|  \Big(\oint_{\p R} \frac{d w}{2\pi i} \bar D^{\bar\alpha}_{\dot A}(w,\bar w)\Big)\bar D^{\bar\beta}_{\dot B}(w_{1},\bar w_{1}) \Big|O^{(0)}_{a}\left(-\frac{T}{2}\right)\Big\rangle\nn
&=&\frac{1}{\pi}\int d^{2}w_{1} \Big\langle O^{(0)}_{b}\left(\frac{T}{2}\right)\Big| \Big( \oint_{|w-w_{1}|=\epsilon} \frac{d w}{2\pi i} \bar D^{\bar\alpha}_{\dot A}(w,\bar w)\Big)
\bar D^{\bar\beta}_{\dot B}(w_{1},\bar w_{1}) \Big|O^{(0)}_{a}\left(-\frac{T}{2}\right)\Big\rangle\nn
&&-\frac{1}{\pi}\int d^{2}w_{1} \Big\langle O^{(0)}_{b}\left(\frac{T}{2}\right)\Big| \Big( \oint_{C_{\frac{T}{2}-\epsilon''}} \frac{d w}{2\pi i} \bar D^{\bar\alpha}_{\dot A}(w,\bar w)\Big)
\bar D^{\bar\beta}_{\dot B}(w_{1},\bar w_{1}) \Big|O^{(0)}_{a}\left(-\frac{T}{2}\right)\Big\rangle\nn
&&+\frac{1}{\pi}\int d^{2}w_{1} \Big\langle O^{(0)}_{b}\left(\frac{T}{2}\right)\Big|  \bar D^{\bar\beta}_{\dot B}(w_{1},\bar w_{1}) 
\Big(\oint_{C_{-\frac{T}{2}+\epsilon''}} \frac{d w}{2\pi i} \bar D^{\bar\alpha}_{\dot A}(w,\bar w)\Big)\Big|O^{(0)}_{a}\left(-\frac{T}{2}\right)\Big\rangle\nn
\eea
We will now evaluate the three terms in the final expression of this equation. 

\subsection{First term of $(\tilde X^{\bar\alpha\bar\beta}_{\dot A\dot B})_{ba}(T)$ in (\ref{GN3terms})}

The OPE between two operators $\bar D$ is derived in  Appendix \ref{DD OPE}. Using this OPE, we find 
\be
\oint_{|w-w_{1}|=\epsilon}\frac{dw}{2\pi i}\bar D^{\bar\alpha}_{\dot A}(w,\bar w)\bar D^{\bar\beta}_{\dot B}(w_1,\bar w_1)= \epsilon_{\dot A \dot B}\epsilon^{\bar\alpha\bar\beta}\pi^2\oint_{|w-w_{1}|=\epsilon}\frac{dw}{2\pi i} \frac{1}{(w-w_1)^2}\frac{1}{\bar w-\bar w_1}
= \epsilon_{\dot A \dot B}\epsilon^{\bar\alpha\bar\beta} \frac{\pi^2}{\epsilon^2}\\
\ee
The regular terms in the $\bar D^{\bar\alpha}_{\dot A}\bar D^{\bar\beta}_{\dot B}$ OPE give a vanishing contribution in the limit $\epsilon\rightarrow 0$.
Then the first term in the last expression in (\ref{GN3terms}) becomes
\be\label{term1 final}
\epsilon_{\dot A \dot B} \epsilon^{\bar\alpha\bar\beta} \frac{\pi}{\epsilon^2}\int d^{2}w \Big\langle O^{(0)}_{b}\left(\frac{T}{2}\right)   \Big|O^{(0)}_{a}\left(-\frac{T}{2}\right)\Big\rangle
=\epsilon_{\dot A \dot B} \epsilon^{\bar\alpha\bar\beta} \frac{2\pi^2 T}{\epsilon^2}e^{-E^{(0)}T} \Big\langle O^{(0)}_{b}   \Big|O^{(0)}_{a}\Big\rangle
\ee
Note that the  large $T$ behavior of this term has the form $\sim Te^{-E^{(0)}T}$, so it will contribute to the lift of conformal dimensions.

\subsection{Second term of $(\tilde X^{\bar\alpha\bar\beta}_{\dot A\dot B})_{ba}(T)$ in (\ref{GN3terms})}

Now consider  the second term in the last expression in (\ref{GN3terms}). In the contour integral we write  $dw=id\sigma$. The term then becomes
\be\label{second term}
-\frac{1}{\pi}\int d^{2}w_{1} \Big\langle O^{(0)}_{b}\left(\frac{T}{2}\right)\Big|
\left(\int^{2\pi}_{0} \frac{d \sigma}{2\pi} \bar D^{\bar\alpha}_{\dot A}(\frac{T}{2}-\epsilon'',\sigma)\right)
\bar D^{\bar\beta}_{\dot B}(w_{1},\bar w_{1}) 
\Big|O^{(0)}_{a}\left(-\frac{T}{2}\right)\Big\rangle
\ee

Consider the states $|\chi\rangle$ that run between the the two $\bar D$ operators; i.e., the states in the region $\tau_1<\tau_\chi<{T\over 2}-\epsilon''$. We will first argue that these states  have  energy $E_{\chi}\geq E^{(0)}$, and then that only the states with $E_{\chi}= E^{(0)}$
will give a contribution relevant to the large $T$ limit that we need. We will then apply a projection operator ${\mathcal P}$ to project onto the states with $E_{\chi}= E^{(0)}$.

Since we have an integral $\int d\sigma$ over the insertion of  $\bar D^{\bar\alpha}_{\dot A}$,
the states $|\chi\rangle$   must have the same spin as the $O^{(0)}_{b}$; i.e.  $h_{\chi}-\bar h_{\chi}=h_{b}-\bar h_{b}$. Because $\bar h_{b}$ has the  lowest allowed dimension for the right movers, we have $\bar h_{\chi}\geq \bar h_{b}$. This gives  $ h_{\chi}\geq  h_{b}$ and
\be
E_{\chi}\geq E^{(0)}
\ee
For the case $E_{\chi}= E^{(0)}$,  time evolution gives the  factor $e^{-E^{(0)}T}$ for all values of $w_1$. The integral of $\int d^2 w_1$ thus gives a factor $T-2\epsilon''$.  In the limit $T\r\infty$ we can write $T-2\epsilon''\r T$,   and we will drop the regulator $\epsilon''$ from now  on. Thus we see that the states with $E_{\chi}= E^{(0)}$ give the behavior $\sim T e^{-E^{(0)}T}$ required for a nonvanishing contribution to the lift of dimensions. 

Now consider states $|\chi\rangle$  with $E_{\chi}> E^{(0)}$.  From the region $\tau_1<\tau_{\chi}<\frac{T}{2}$, the time evolution gives a factor $e^{-E_{\chi}(\frac{T}{2}-\tau_1)}$. From the region $-\frac{T}{2}<\tau_{\chi}<\tau_1$ the time evolution gives a factor $e^{-E^{(0)}(\tau_1+\frac{T}{2})}$. Thus the integral $\int d^{2}w_{1}$ gives a $T$ dependence at large $T$
\be
\int_{-T/2}^{T/2}d\tau_1 e^{-E_{\chi}(\frac{T}{2}-\tau_1)}e^{-E^{(0)}(\tau_1+\frac{T}{2})}
=\frac{e^{-E^{(0)}T}-e^{-E_{\chi}T}}{E_{\chi}-E^{(0)}}
\rightarrow \frac{1}{E_{\chi}-E^{(0)}} e^{-E^{(0)}T}
\ee
We see that states with $E_{\chi}\geq E^{(0)}$ give a large $T$ contribution of the form $\sim  e^{-E^{(0)}T}$ instead of the required behavior $ T e^{-E^{(0)}T}$.

Thus we need to only consider   intermediate states $|\chi\rangle$ with 
\be\label{Pcon}
E_{\chi}= E^{(0)}
\ee
Let  $\mathcal{P}$ be the projection operator that projects the state at any time $\tau$ onto the subspace (\ref{Pcon}).  Taking into account only intermediate states $|\chi\rangle$ satisfying (\ref{Pcon}) allows us to write 
\be
\int d^{2}w \r 2\pi T\int_{0}^{2\pi} \frac{d\sigma}{2\pi}
\ee
Then the second term in the last expression  in eq. (\ref{GN3terms}) becomes
\be\label{GN2}
-2 T \Big\langle O^{(0)}_{b}\left(\frac{T}{2}\right)\Big| 
\left(\int_{0}^{2\pi} \frac{d \sigma}{2\pi} \bar D^{\bar\alpha}_{\dot A}(\frac{T}{2},\sigma)\right)
\mathcal P 
\left(\int_{0}^{2\pi} \frac{d\sigma}{2\pi} \bar D^{\bar\beta}_{\dot B}(\tau,\sigma)\right) \Big|O^{(0)}_{a}\left(-\frac{T}{2}\right)\Big\rangle
\ee

Our next task is to bring all states and operators to the same time, say  $\tau=0$. First note that all intermediate states in (\ref{GN2}) have the same dimension, so we can slide the 
operators $\bar D^{\bar\beta}_{\dot B}$ and  $\bar D^{\bar\alpha}_{\dot A}$  freely up and down the cylinder as long as we do not cross another operator or state insertion. Thus (\ref{GN2}) can be written as
\be
-2 T e^{-E^{(0)}T} \Big\langle O^{(0)}_{b}\left(0\right)\Big|  
\left(\mathcal P \int_{0}^{2\pi} \frac{d\sigma'}{2\pi} \bar D^{\bar\alpha}_{\dot A}(0,\sigma') \mathcal P\right)
\left(\mathcal P\int_{0}^{2\pi} \frac{d\sigma}{2\pi} \bar D^{\bar\beta}_{\dot B}(0,\sigma)  \mathcal P \right)
\Big| O^{(0)}_{a}\left(0\right)\Big\rangle\nn
\ee
where we add some extra projection operators; this can be done using ${\mathcal P}^2={\mathcal P}$ and the fact the initial and final states $O_a$ and $O_b$ have the energy $E^{(0)}$ (\ref{Pcon}).
Now we define the following operator
\be\label{GP}
\bar G^{\bar\alpha (P)}_{\dot A,0}= \mathcal P\int_{0}^{2\pi} \frac{d\sigma}{2\pi} \bar D^{\bar\alpha}_{\dot A}(\tau,\sigma)  ~ \mathcal P
\ee
It is an operator independent of $\tau$: we can slide the contour up and down the cylinder as long as it does not intersect another operator insertion. Using the above defined operator, the (\ref{second term}) can be written as
\be\label{term2 final}
-2 T e^{-E^{(0)}T}\Big\langle O^{(0)}_{b}\Big|\bar G^{\bar\alpha(P)}_{\dot A,0} \bar G^{\bar\beta(P)}_{\dot B,0} \Big| O^{(0)}_{a}\Big\rangle
\ee
Here all the operators and states are at time $\tau=0$.

\subsection{Third term of $(\tilde X^{\bar\alpha\bar\beta}_{\dot A\dot B})_{ba}(T)$ in (\ref{GN3terms})}

The third term in the  last expression in (\ref{GN3terms}) is handled in the same way as the second term.  We write he contour as $ d w= -i d\sigma$ at the boundary $-\frac{T}{2}$. This gives 
\bea \label{term3 final}
&&-\frac{1}{\pi} \Big\langle O^{(0)}_{b}\left(\frac{T}{2}\right)\Big|  
\left(\int d^2 w_{1}\bar D^{\bar\beta}_{\dot B}(w_{1},\bar w_{1})  \right)
\left(\int^{2\pi}_{0} \frac{d\sigma}{2\pi} \bar D^{\bar\alpha}_{\dot A}(-\frac{T}{2},\sigma) \right)
\Big|O^{(0)}_{a}\left(-\frac{T}{2}\right)\Big\rangle\nn
&\rightarrow&-2 T \Big\langle O^{(0)}_{b}\left(\frac{T}{2}\right)\Big|  
\left(\int_{0}^{2\pi} \frac{d\sigma}{2\pi}\bar D^{\bar\beta}_{\dot B}(\tau,\sigma)  \right)
\mathcal P 
\left(\int^{2\pi}_{0} \frac{d\sigma'}{2\pi} \bar D^{\bar\alpha}_{\dot A}(\tau',\sigma') \right)
\Big|O^{(0)}_{a}\left(-\frac{T}{2}\right)\Big\rangle\nn
&=&-2 T e^{-E^{(0)}T}\Big\langle O^{(0)}_{b}\Big| \bar G^{\bar\beta(P)}_{\dot B,0}\bar G^{\bar\alpha(P)}_{\dot A,0} \Big| O^{(0)}_{a}\Big\rangle
\eea

\subsection{Final expression for the large $T$ behavior of $(\tilde X^{\bar\alpha\bar\beta}_{\dot A\dot B})_{ba}(T)$}

We now sum over the three terms (\ref{term1 final}), (\ref{term2 final}) and (\ref{term3 final}). We find that  the large $T$ behavior of amplitude $\tilde X(T)$ (\ref{Xamp}) becomes
\be\label{X large T}
\lim_{T\rightarrow \infty} (\tilde X^{\bar\alpha\bar\beta}_{\dot A\dot B})_{ba}(T)=
2 T e^{-E^{(0)}T}   
\Big\langle O^{(0)}_{b}\Big|    -\Big\{  \bar G^{\bar\alpha(P)}_{\dot A,0},  \bar G^{\bar\beta(P)}_{\dot B,0} \Big\} +\epsilon_{\dot A \dot B} \epsilon^{\bar\alpha\bar\beta} \frac{\pi^2}{\epsilon^2}   \Big|O^{(0)}_{a}\Big\rangle
\ee
where
\be\label{def tilde X}
(\tilde X^{\bar\alpha\bar\beta}_{\dot A\dot B})_{ba}(T)= \Big\langle O^{(0)}_{b}\left(\frac{T}{2}\right)\Big| 
\left(\int d^{2}w_1  \bar G^{\bar\alpha(0)}_{\dot A,-\frac{1}{2}}\bar D^{\bar\beta}_{\dot B}(w_1,\bar w_1)\right)
\left(\int d^{2}w_{2}D(w_{2},\bar w_{2}) \right)
\Big|O^{(0)}_{a}\left(-\frac{T}{2}\right)\Big\rangle\\
\ee
In this large T limit, the only regulator left is one setting the minimal distance $\epsilon$ between $w_{1}$ and $w_{2}$
\be
|w_{1}-w_{2}|>\epsilon
\ee

\section{Deriving the Gava-Narain relation}\label{sec 6}

We are interested in the lifting of states that are half-supersymmetric at the orbifold pont; i.e., states that are annihilated by the right supercharges before perturbation by the deformation $D$. We noted that the first order lift vanished for such states, and that the second order lift was given through the matrix
\be\label{lift matrix}
E^{(2)}_{ba}=\lim_{T\rightarrow \infty}-\frac{\lambda^2}{2T}e^{E^{(0)}T}X_{ba}(T)+ \frac{2\pi^2 \lambda^2}{\epsilon^2}\delta_{ba}
\ee
where
\be\label{Xamp}
X_{ba}(T)=\Big\langle O^{(0)}_{b}\left(\frac{T}{2}\right)\Big|\left(\int d^{2}w_{1}D(w_{1},\bar w_{1})\right)\left(\int d^{2}w_{2}D(w_{2},\bar w_{2})\right)\Big|O^{(0)}_{a}\left(-\frac{T}{2}\right)\Big\rangle
\ee
Here the range of the $w_{i}$ integrals are given in section \ref{sec 5}.
An expression for  $D(w_1,\bar w_1)$ was given in  (\ref{deformation 1}). This expression has two terms. Each of these terms can be written as an expression of the type $\tilde X^{\bar\alpha\bar\beta}_{\dot A\dot B}(T)$ defined in (\ref{def tilde X}). We find that the amplitude $X_{ba}(T)$ can be written as
\be
X(T)=-\tilde X^{-+}_{-+}(T)-\tilde X^{+-}_{+-}(T)
\ee
where we have suppressed the $a,b$ indices. 

We now use our evaluation of the $\tilde X^{\bar\alpha\bar\beta}_{\dot A\dot B}(T)$, in the limit $T\r\infty$. Using eq. (\ref{X large T}), we find 
\bea
\lim_{T\rightarrow \infty}  X_{ba}(T)&=&
4 T e^{-E^{(0)}T}   
\Big\langle O^{(0)}_{b}\Big|    \Big\{  \bar G^{-(P)}_{-,0},  \bar G^{+(P)}_{+,0} \Big\}  +\frac{\pi^2}{\epsilon^2}  \Big|O^{(0)}_{a}\Big\rangle
\eea
We see the expected large $T$ behavior  $T e^{-E^{(0)}T}$ which gives the contribution to  the lifting matrix (\ref{lift matrix}). The lifting matrix becomes
\be\label{GN final 1}
E^{(2)}_{ba}=-2 \lambda^2   \Big\langle O^{(0)}_{b}\Big|    \Big\{  \bar G^{-(P)}_{-,0},  \bar G^{+(P)}_{+,0} \Big\}  \Big|O^{(0)}_{a}\Big\rangle
\ee

Similarly, express the  $D(w_1,\bar w_1)$ in terms of (\ref{deformation 2}), the amplitude $X_{ba}(T)$ can be written as
\be
X(T)=\tilde X^{-+}_{+-}(T)+\tilde X^{+-}_{-+}(T)
\ee
Using eq. (\ref{X large T}), the large T behavior is
\bea
\lim_{T\rightarrow \infty}  X_{ba}(T)&=&
-4 T e^{-E^{(0)}T}   
\Big\langle O^{(0)}_{b}\Big| 
\Big\{   \bar G^{-(P)}_{+,0} ,\bar G^{+(P)}_{-,0} \Big\}  -\frac{\pi^2}{\epsilon^2}  \Big|O^{(0)}_{a}\Big\rangle
\eea
As we can see, this indeed has the expected large $T$ behavior to contribute to the lifting matrix (\ref{lift matrix}). 
The lifting matrix becomes
\be\label{GN final 2}
E^{(2)}_{ba}=2 \lambda^2   \Big\langle O^{(0)}_{b}\Big| 
\Big\{   \bar G^{-(P)}_{+,0},\bar G^{+(P)}_{-,0} \Big\} 
\Big|O^{(0)}_{a}\Big\rangle
\ee

Let us now use the Hermitian conjugation relation derived in  Appendix \ref{app_cft}
\be
\bar G^{+(P)\dagger}_{+,0}=-\bar G^{-(P)}_{-,0}~~~~~~
\bar G^{+(P)\dagger}_{-,0}=\bar G^{-(P)}_{+,0}
\ee
Then we  get the two equivalent expressions for the lifting matrix,  from (\ref{GN final 1}) and (\ref{GN final 2}) respectively:
\be\label{GNe1}
E^{(2)}_{ba}=2 \lambda^2   
\Big\langle O^{(0)}_{b}\Big|    \Big\{  \bar G^{+(P)\dagger}_{+,0},  \bar G^{+(P)}_{+,0} \Big\} \Big|O^{(0)}_{a}\Big\rangle
\ee
and
\be\label{GNe2}
E^{(2)}_{ba}=2 \lambda^2   
\Big\langle O^{(0)}_{b}\Big|    \Big\{  \bar G^{+(P)\dagger}_{-,0},  \bar G^{+(P)}_{-,0} \Big\} \Big|O^{(0)}_{a}\Big\rangle
\ee

It will be convenient to define an operator
\be\label{lifting operator}
\hat E^{(2)}=2 \lambda^2   
  \Big\{  \bar G^{+(P)\dagger}_{+,0},  \bar G^{+(P)}_{+,0} \Big\} = 2 \lambda^2   
    \Big\{  \bar G^{+(P)\dagger}_{-,0},  \bar G^{+(P)}_{-,0} \Big\} 
\ee
Evaluating this operator between the states $\langle O^{(0)}_b|$ and $|O^{(0)}_a\rangle$ gives the lifting matrix element  $E^{(2)}_{ba}$.

\section{The structure of long super-multiplets}\label{sec super multi}

States which are annihilated by some supercharges fall into short multiplets. States which are not annihilated by supercharges fall into long multiplets.

Each state in a multiplet should have the same energy. This follows from the supersymmetry algebra: the supercharges commute with the Hamiltonian. 

Our goal is to compute the lift of energies when the states are perturbed off the orbifold point. At the orbifold point itself, the relevant supercharges are given by the zeroth order operators $\bar G^{\bar \alpha(0)}_{\dot A, 0}$. The states $|O^{(0)}_a\rangle$ we consider are annihilated by all these supercharges, so they fall into short multiplets at the orbifold point.

Under perturbation, the energy can lift. But this requires that  the short multiplets join to longer multiplets. The lift must be the  same for all other members of the long multiplet. Thus if compute the lift for one member of the multiplet, we should not need to compute it for other members of the multiplet. 

But how do we find the other members of the multiplet? If we know the supercharge, then applying this to the state should take us to other members of the multiplet. Since we are working to $O(\lambda^2)$, we need to know in principle the operator $\bar G=\bar G^{(0)}+\lambda \bar G^{(1)}+\lambda^2 \bar G^{(2)}$. But we do not know this operator. We have instead defined operators $\bar G^{\bar \alpha(0)}_{\dot A, 0}$ that work operate only in the subspace of states $|O^{(0)}_a\rangle$. We have shown that the anticommutator of these operators gives the lift of states of type $|O^{(0)}_a\rangle$. It is not however a priori clear that we can use the  $\bar G^{\bar \alpha(0)}_{\dot A, 0}$ to take us to other states in the multiplet. 

We will now check explicitly that the $\bar G^{\bar \alpha(0)}_{\dot A, 0}$ have the properties required of the operators that move us up and down a long supermultiplet.

Consider the following two operators, which we will call the `raising' operators of the supermultiplet:
\be
\bar G^{+(P)}_{-,0},~~~~~\bar G^{+(P)}_{+,0}
\ee
First we will check that 
\be
\{\bar G^{+(P)}_{\dot A,0},\bar G^{+(P)}_{\dot B,0}\}=0
\ee
This condition is of course necessary if these operators are to generate a supermultiplet. Next we  will check that  the four states obtained by application of these operators have the same lift. This is equivalent to the condition
\be
[\hat E^{(2)}, \bar G^{+(P)}_{\dot A,0}]=0
\ee

\subsection{Showing $\{\bar G^{+(P)}_{\dot A,0},\bar G^{+(P)}_{\dot B,0}\}=0$}

Consider the following amplitude
\be
(\tilde X^{++}_{\dot A\dot B})_{ba}(T)= \Big\langle O^{(0)}_{b}\left(\frac{T}{2}\right)\Big| \int d^{2}w_1  \bar G^{+}_{\dot A,-\frac{1}{2}}\bar D^{+}_{\dot B}(w_1,\bar w_1)\int d^{2}w_{2}D(w_{2},\bar w_{2}) \Big|O^{(0)}_{a}\left(-\frac{T}{2}\right)\Big\rangle
\ee
Since 
\be
  \bar G^{+(0)}_{\dot A,-\frac{1}{2}}\bar D^{+}_{\dot B}=\bar G^{+(0)}_{\dot A,-\frac{1}{2}} G^{+(0)}_{\dot B,-\frac{1}{2}}\sigma^{-+}=0
\ee
we have
\be\label{X++=0}
(\tilde X^{++}_{\dot A\dot B})_{ba}(T)=0
\ee
The large $T$ behavior (\ref{X large T}) is
\be
\lim_{T\rightarrow \infty} (\tilde X^{++}_{\dot A\dot B})_{ba}(T)=
-2 T e^{-E^{(0)}T}   
\Big\langle O^{(0)}_{b}\Big|    \Big\{  \bar G^{+(P)}_{\dot A,0},  \bar G^{+(P)}_{\dot B,0} \Big\}  \Big|O^{(0)}_{a}\Big\rangle
\ee
Thus from eq. (\ref{X++=0}), we have 
\be\label{PG fermion}
\{\bar G^{+(P)}_{\dot A,0},\bar G^{+(P)}_{\dot B,0}\}=0
\ee

\subsection{Equal lifting for states related by $\bar G^{\bar \alpha(P)}_{\dot A}$}

We have
\bea
&&\left[\Big\{  \bar G^{+(P)\dagger}_{+,0},  \bar G^{+(P)}_{+,0} \Big\}, \bar G^{+(P)}_{+,0}\right]\nn
&=&\hspace{-2 pt}\left(\bar G^{+(P)\dagger}_{+,0} \bar G^{+(P)}_{+,0}+\bar G^{+(P)}_{+,0} \bar G^{+(P)\dagger}_{+,0}\right)\bar G^{+(P)}_{+,0}
-\bar G^{+(P)}_{+,0}\left(\bar G^{+(P)\dagger}_{+,0} \bar G^{+(P)}_{+,0}+\bar G^{+(P)}_{+,0} \bar G^{+(P)\dagger}_{+,0}\right)\nn
\eea
Using the relation $\bar G^{+(P)}_{+,0}\bar G^{+(P)}_{+,0}=0$, which we get from eq. (\ref{PG fermion}), we find 
\be
\left[\Big\{  \bar G^{+(P)\dagger}_{+,0},  \bar G^{+(P)}_{+,0} \Big\}, \bar G^{+(P)}_{+,0}\right]=0
\ee
Then from (\ref{lifting operator}) we have
\be
[\hat E^{(2)}, \bar G^{+(P)}_{+,0}]=0
\label{vtwone}
\ee
Similarly, from 
\be
\left[\Big\{  \bar G^{+(P)\dagger}_{-,0},  \bar G^{+(P)}_{-,0} \Big\}, \bar G^{+(P)}_{-,0}\right]=0
\ee
we find
\be
[\hat E^{(2)}, \bar G^{+(P)}_{-,0}]=0
\label{vtwtwo}
\ee

Combining the relations (\ref{vtwone}),(\ref{vtwtwo}), we can write
\be
[\hat E^{(2)}, \bar G^{+(P)}_{\dot A,0}]=0
\ee
Thus we see that states that are related by the application of $\bar G^{\bar \alpha(P)}_{\dot A}$ have the same lift. We can therefore use these operators to group the states into supermultiplets, all of whose members have the same lift.

\section{Deriving the lifting in the Hamiltonian formalism}\label{sec 8}

Recall that our main goal was to understand how the lift could be obtained by a simple expression like (\ref{zzten}), while a direct expansion in terms of states and operators gives a more complicated expression given by (\ref{Xi}),(\ref{vten}). The simpler expression (\ref{zzten}) needs only a first order computation; the second order lift is obtained, roughly speaking,  from squaring this first order computation. The more complicated expression (\ref{vten}) involves a first order correction to the state $|O^{(1)}\rangle$, as well as a first order correction to the supercharge $\bar G^{(1)}$. 

In the preceding sections we have seen how starting from the path integral at second order in perturbation theory, we can obtain the simple result (\ref{zzten}) (or rather, its extension to the more general situation where we have operator mixing). In this section we will see that the more complicated expression (\ref{vten}) reduces to the simple expression (\ref{zzten}) after we understand the structure of $|O^{(1)}\rangle$  and $\bar G^{(1)}$. Thus the discussion below will serve to clarify the relation between the various objects we have encountered; in particular, the operators $\bar G^{(1)}, \bar G^{(P)}$. 

\subsection{The first order correction  $|O^{(1)}\rangle$ to the state}

In Hamiltonian  perturbation theory, the first order correction to an eigenstate is given by
\be
|O^{(1)}\rangle
=-\sum_{k\neq 0}\frac{\langle O^{(0)}_{k}| H^{(1)} |O^{(0)}\rangle}{E_{k}-E^{(0)}}\, |O^{(0)}_{k}\rangle
\label{vthone}
\ee
This expression is valid for nondegenerate perturbation theory. In our case we do have in general several states with the same energy as  $|O\rangle$. But as we will see below, $H^{(1)}$ has a vanishing matrix element between these degenerate states. In this situation (\ref{vthone}) is valid again. 

We now wish to find an expression for $|O^{(1)}\rangle$ in the path integral language that we have been using. First we obtain the coefficients in $|O^{(1)}\rangle$ by considering the following amplitude 
\be
\lim_{T\rightarrow \infty}\langle O^{(0)}_{k}|e^{-(H^{(0)}+\lambda H^{(1)})T} |O^{(0)}\rangle
\ee
The order $\lambda$ contribution  is
\bea
{\cal A}&=&-\lim_{T\rightarrow \infty}\int^{T}_{0} dt \langle O^{(0)}_{k}| e^{-H^{(0)}(T-t)}H^{(1)}e^{-H^{(0)}t} |O^{(0)}\rangle\nn
&=&-\lim_{T\rightarrow \infty}\int^{T}_{0} dt e^{-E_{k}(T-t)}e^{-E^{(0)}t}\langle O^{(0)}_{k}| H^{(1)} |O^{(0)}\rangle\nn
&=&- \lim_{T\rightarrow \infty}\frac{e^{-E^{(0)}T}-e^{-E_{k}T}}{E_{k}-E^{(0)}}\langle O^{(0)}_{k}| H^{(1)} |O^{(0)}\rangle\nn
&=&- {e^{-E^{(0)}T}}\frac{\langle O^{(0)}_{k}| H^{(1)} |O^{(0)}\rangle}{E_{k}-E^{(0)}}
\label{vthfive}
\eea
Thus we see that the coefficients in (\ref{vthone}) can be obtained in terms of  amplitudes of the form  ${\cal A}$.  

Now we write ${\cal A}$ in a path integral language. Since ${\cal A}$ computes the transition amplitude to first order in perturbation theory, we find
\be\label{LF}
{\cal A}=\lim_{T\rightarrow \infty}\langle O^{(0)}_{k}(0)|\int d^{2}w D(w,\bar w) |O^{(0)}(-T)\rangle
\ee
Here the zeroth order term vanishes for  $|O^{(0)}_k\rangle \ne |O^{(0)}\rangle$.

We have taken the domain of integration of the $D$ to be bounded by curves $\tau=constant$ rather than curves of any other shape. In Appendix \ref{appd} we argue that this prescription is needed for the validity of the derivation in \cite{hmz} of the expression (\ref{moti h matrix1}) for the lifting $\delta^{(2)} \bar h$. With this choice, we find that $|O^{(0)}_k\rangle$ has energy $E_k>E^{(0)}$, for the following reason. 
Using (\ref{deformation 1}) we can rewrite the deformation operator schematically as 
\be\label{deformation Hami}
D(w,\bar w)\sim \oint_{C_{w}} d\bar w' \bar G^{(0)}(\bar w') \bar D(w,\bar w)\sim  \Big\{\bar G^{(0)}_{,0}~,~ \bar D(w,\bar w)\Big\}
\ee
The $\bar G^{(0)}_{,0}$ kills  right moving Ramond ground states.  If the state $|O^{(0)}_k\rangle$ is a right moving Ramond ground state, both terms in the anticommutator (\ref{deformation Hami}) vanish.
Thus we get $E_k>E^{(0)}$. This also tells us that $|O^{(0)}_k\rangle$ in  (\ref{vthone}) has $E_k>E^{(0)}$, thus allowing us to use the expression (\ref{vthone}). 
 
Equating (\ref{vthfive}) and (\ref{LF}), we get
\be
\lim_{T\rightarrow \infty}\langle O^{(0)}_{k}(0)|\int d^{2}w D(w,\bar w) |O^{(0)}(-T)\rangle=- \frac{\langle O^{(0)}_{k}| H^{(1)} |O^{(0)}\rangle}{E_{k}-E^{(0)}}
e^{-E^{(0)}T}
\ee
Using (\ref{vthone}) we finally get
\be
|O^{(1)}\rangle
=\lim_{T\rightarrow \infty}e^{E^{(0)}T}\int d^{2}w D(w,\bar w) |O^{(0)}(-T)\rangle 
\label{vthseven}
\ee
Thus we see that the first order correction $|O^{(1)}\rangle$ to the state can be obtained, as a state at $\tau=0$ as follows. We start with the leading order state $|O^{(0)}\rangle$ at $T\r-\infty$, do an integral over the deformation $D$ in the region $-T<\tau<0$, and rescale by the factor $e^{E^{(0)}T}$.

\subsection{Construction of $\bar G^{\bar\alpha(0)}_{\dot A,0}|O^{(1)}\rangle$}

Having constructed a path integral expression (\ref{vthseven}) for $|O^{(1)}\rangle$, we now wish to construct a similar expression for the leading order supercharge acting on $|O^{(1)}\rangle$. We have
\be
\bar G^{\bar\alpha(0)}_{\dot A,0}|O^{(1)}\rangle
=-\lim_{T\rightarrow \infty}\oint_{C_{\tau=0}} \frac{d\bar w'}{2\pi i} \bar G^{\bar\alpha(0)}_{\dot A} (w',\bar w')
\int d^{2}w D(w,\bar w) |O^{(0)}(-T)\rangle e^{E^{(0)}T}
\label{vtheight}
\ee
We can write this in a way where the $\bar G^{\bar\alpha(0)}_{\dot A}$ integral surround the entire region where $D$ is integrated. This can be done by noting that $\bar G^{\bar\alpha(0)}_{\dot A}$ annihilates the right mover Ramond ground state $|O^{(0)}\rangle$. So we can add to (\ref{vtheight}) the vanishing expression
\be
0=\lim_{T\rightarrow \infty}
\int d^{2}w D(w,\bar w) \oint_{C_{\tau=\infty}} \frac{d\bar w'}{2\pi i} \bar G^{\bar\alpha(0)}_{\dot A} (w',\bar w') |O^{(0)}(-T)\rangle e^{E^{(0)}T}
\ee
We can now apply the general result (\ref{G to GN}), to get
\bea
&&\bar G^{\bar\alpha(0)}_{\dot A,0}|O^{(1)}\rangle \nn
&=& \hspace{-4 pt}- \lim_{T\rightarrow \infty}\left[ \oint_{C_{\tau=0}} \frac{d w'}{2\pi i} \bar D^{\bar\alpha}_{\dot A} (w',\bar w')
|O^{(0)}(-T)\rangle +\oint_{C_{\tau=-T}} \frac{d w'}{2\pi i} \bar D^{\bar\alpha}_{\dot A} (w',\bar w')
|O^{(0)}(-T)\rangle\right]e^{E^{(0)}T}\nn
&=& \hspace{-4 pt}- \lim_{T\rightarrow \infty}\left[ \int^{2\pi}_{0} \frac{d \sigma}{2\pi } 
\bar D^{\bar\alpha}_{\dot A} (\tau=0,\sigma)
|O^{(0)}(-T)\rangle -\int^{2\pi}_{0} \frac{d \sigma}{2\pi } \bar D^{\bar\alpha}_{\dot A} (\tau=-T, \sigma)
|O^{(0)}(-T)\rangle\right]e^{E^{(0)}T}\nn
\eea
where we use $dw=id\sigma$ at $\tau=0$ and $dw=-id\sigma$ at $\tau=-T$. 
In the first term, since the operator $\bar D^{\bar\alpha}_{\dot A}$ is placed at $\tau=0$, we have
$|O^{(0)}(-T)\rangle e^{E^{(0)}T}=|O^{(0)}\rangle$. The second term gives the $\bar G^{P}$ operator in the large $T$ limit. Thus
\be
\bar G^{\bar\alpha(0)}_{\dot A,0}|O^{(1)}\rangle
=-\bar G^{\bar\alpha(1)}_{\dot A,0}|O^{(0)}\rangle
+ \bar G^{\bar\alpha(P)}_{\dot A,0}|O^{(0)}\rangle
\ee
or
\be
 \bar G^{\bar\alpha(P)}_{\dot A,0}|O^{(0)}\rangle=
 \bar G^{\bar\alpha(1)}_{\dot A,0}|O^{(0)}\rangle+
 \bar G^{\bar\alpha(0)}_{\dot A,0}|O^{(1)}\rangle
\ee
This is the relation which explains the difference between the simple expression (\ref{zzten}) obtained by Gava-Narain where we compute a finite number of 3 point functions, and the more complicated expression (\ref{vten}) which resolve into an infinite number of 3-point functions.

\section{Discussion}\label{discussion}

The D1D5 CFT is an important model for black holes in string theory. While this system is easy to study at its `free point' -- the orbifold point -- one needs to deform away from this point to understand the interactions that describe the gravity theory of interest. The deformation is given by a perturbation operator $D$, but finding amplitudes with insertions of $D$ is complicated as $D$ contains a second order twist. 

We are interested in states which are supersymmetric at the orbifold point but which might lift to higher energies at second order in the deformation parameter $\lambda$. This lift can be computed by a  path-integral of the type (\ref{moti h matrix1}) containing two insertions $D$ whose positions need to be integrated over. The states at $\tau=\pm\infty$ are usually taken to be the leading order states $|O^{(0)}_a\rangle$, and it was shown in \cite{hmz} that we can indeed ignore the corrections $|O^{(1)}_a\rangle, |O^{(2)}_a\rangle$ in the path integral formulation when computing the lift of states that are supersymmetric at the orbifold point.

The amplitude in the path integral is a 4-point function. In the examples of such computations  in \cite{gz, hmz} the right-moving part of the amplitude could be explicitly computed; this is the case whenever the covering space obtained by undoing the twists is a sphere. But we also encounter the situation where the covering space is a torus, and then it is very difficult to find the needed 4-point function explicitly. For this situation we would therefore like to use an expression for the lift proposed by Gava and Narain \cite{gn} which required the computation of only a finite number of 3-point functions.  The 3-point functions can be computed readily since the covering space in these cases is a sphere. 

In understanding the formulation of \cite{gn} we again come across the issue of whether we need the corrections $|O^{(1)}_a\rangle$ to the state. The expression they propose needs only the leading order part $|O^{(0)}_a\rangle$. But a direct Hamiltonian derivation using the superconformal algebra does appear to bring in $|O^{(1)}_a\rangle$. In this derivation, carried out using the expansion (\ref{zzel}),  one finds the operator $\bar G^{\bar\alpha(1)}_{\dot A}$ as the first order correction to the supercharge. Acting with $\bar G^{\bar\alpha(1)}_{\dot A}$ on $|O^{(0)}_a\rangle$ generates an infinite number of states with arbitrarily high energies. By contrast,   in the prescription of \cite{gn} we get only a finite number of states with energy equalling the energy of the $|O^{(0)}_a\rangle$. Thus we seem to have two different looking expressions for the lift: eq. (\ref{vten}) from a direct Hamiltonian formulation and a simpler expression (\ref{zzten}) from the proposal of \cite{gn}. 

We relate these two different looking results through the expression (\ref{zzthir}).  A key point is that  the first order correction to the supercharge used in \cite{gn} is not $\bar G^{\bar\alpha(1)}_{\dot A}$ but a projection $ \bar G^{\bar\alpha (P)}_{\dot A}$ defined in (\ref{zzfift}). The difference between $\bar G^{\bar\alpha(1)}_{\dot A}$ and $\bar G^{\bar\alpha(P)}_{\dot A}$ compensates for the contribution of the correction  $|O^{(1)}_a\rangle$ found in the Hamiltonian approach. 

We also find that  $\bar G^{\bar\alpha(P)}_{\dot A, 0}$ acts as an operator that can generate  multiplets, all of whose members will have the same value for the lifting. 

We hope to return in future publications to the explicit computation of lifting  for the family of D1D5P states that arise for the extremal black hole in string theory.

 \section*{Acknowledgements}

We would like to thank Philip Argyres, Shaun Hampton and Ida Zadeh for many helpful discussions.  This work is supported in part by DOE grant de-sc0011726.

\pagebreak

\appendix

\section{Notations and conventions} \label{app_cft}

We follow the notation in the appendix of \cite{hmz}. The indices $\alpha=(+,-)$ and $\bar \alpha=(+,-)$ correspond to the subgroup $SU(2)_L$ and $SU(2)_R$ of rotations on $S^3$. The index $\dot A=(+,-)$ corresponds to the subgroup $SU(2)_1$ from rotations in $T^4$. We use the convention
\be
\epsilon_{+-}=1, ~~~\epsilon^{+-}=-1
\ee
The commutators we used in the paper of the right movers are
\bea\label{commutations_1}
\big\{ \bar G^{\bar\alpha}_{\dot{A},-\frac{1}{2}} , \bar G^{\bar\beta}_{\dot{B},-\frac{1}{2}} \big\}&=&  \epsilon_{\dot{A}\dot{B}} \epsilon^{\bar\alpha\bar\beta}\bar L_{-1}  \nn
\big\{ \bar G^{\bar\alpha}_{\dot{A},0} , \bar G^{\bar\beta}_{\dot{B},0} \big\}&=&  \epsilon_{\dot{A}\dot{B}} \epsilon^{\bar\alpha\bar\beta}\left(\bar L_{0}-\frac{ c}{24}\right)
\eea
We have the relations 
\bea\label{A3}
 G_{\dot{A},-\frac{1}{2}}^{+(0)}\sigma^{ + \bar\alpha}=0~~~~~~~~ G_{\dot{A},-\frac{1}{2}}^{-(0)}\sigma^{ -\bar \alpha}=0\nn
\bar G_{\dot{A},-\frac{1}{2}}^{+(0)}\sigma^{ \alpha +}=0~~~~~~~~\bar G_{\dot{A},-\frac{1}{2}}^{-(0)}\sigma^{ \alpha -}=0
\eea
We also have the relations
\be\label{A4}
G_{\dot{A},-\frac{1}{2}}^{-(0)}\sigma^{ + \bar\alpha}=-G_{\dot{A},-\frac{1}{2}}^{+(0)}\sigma^{ - \bar\alpha}
~~~~~~
\bar G_{\dot{A},-\frac{1}{2}}^{-(0)}\sigma^{ \alpha +}=-\bar G_{\dot{A},-\frac{1}{2}}^{+(0)}\sigma^{ \alpha -}
\ee
These can be proved by
\be
G_{\dot{A},-\frac{1}{2}}^{-(0)}\sigma^{ + \bar\alpha}=[J^{-(0)}_{0},G_{\dot{A},-\frac{1}{2}}^{+(0)}]\sigma^{ + \bar\alpha}
=J^{-(0)}_{0}G_{\dot{A},-\frac{1}{2}}^{+(0)}\sigma^{ + \bar\alpha}-G_{\dot{A},-\frac{1}{2}}^{+(0)}J^{-(0)}_{0}\sigma^{ + \bar\alpha}
=-G_{\dot{A},-\frac{1}{2}}^{+(0)}\sigma^{ - \bar\alpha}
\ee
In the last step, we use (\ref{A4}) and $J^{-(0)}_{0}\sigma^{ + \bar\alpha}=\sigma^{ - \bar\alpha}$.
We use the following Hermitian conjugation rule
\bea
( G^{+}_{+}(\tau,\sigma))^{\dagger}=-G^{-}_{-}(-\tau,\sigma)
~~~~~~~
( G^{+}_{-}(\tau,\sigma))^{\dagger}=G^{-}_{+}(-\tau,\sigma)\nn
( \bar G^{+}_{+}(\tau,\sigma))^{\dagger}=-\bar G^{-}_{-}(-\tau,\sigma)
~~~~~~~
( \bar G^{+}_{-}(\tau,\sigma))^{\dagger}=\bar G^{-}_{+}(-\tau,\sigma)
\eea
and 
\be
(\sigma^{--}(\tau,\sigma))^{\dagger}=-\sigma^{++}(-\tau,\sigma)
~~~~~~~
(\sigma^{-+}(\tau,\sigma))^{\dagger}=\sigma^{+-}(-\tau,\sigma)
\ee
In this convention, the deformation operator is a Hermitian operator
\bea
(D(\tau,\sigma))^{\dagger}=D(-\tau,\sigma)
\eea
and for the $\bar D$ operator we have
\be
( \bar D^{+}_{+}(\tau,\sigma))^{\dagger}=-\bar D^{-}_{-}(-\tau,\sigma)
~~~~~~~
( \bar D^{+}_{-}(\tau,\sigma))^{\dagger}=\bar D^{-}_{+}(-\tau,\sigma)
\ee
Thus from the definition of $\bar G^{(P)}$ (\ref{GP}), we have
\be
\bar G^{+(P)\dagger}_{+,0}=-\bar G^{-(P)}_{-,0}~~~~~~
\bar G^{+(P)\dagger}_{-,0}=\bar G^{-(P)}_{+,0}
\ee

\section{ The $\bar D^{\bar\alpha}_{\dot A} \bar D^{\bar\beta}_{\dot B}$ OPE}\label{DD OPE}

In this appendix, we will compute  the OPE 
\be
\bar D^{\bar\alpha}_{\dot A}(w,\bar w)\bar D^{\bar\beta}_{\dot B}(0,0)
\ee
Here the $\bar D$ operator is
\be
\bar D^{\bar\alpha}_{\dot A}= \pi G^{+(0)}_{\dot A}\sigma^{-\bar\alpha}
\ee
The OPE becomes
\bea
&&\bar D^{\bar\alpha}_{\dot A}(w,\bar w)\bar D^{\bar\beta}_{\dot B}(0,0)
=- \pi^2 G^{-(0)}_{\dot A,-\frac{1}{2}}\sigma^{+\bar\alpha}(w,\bar w)G^{+(0)}_{\dot B,-\frac{1}{2}}\sigma^{-\bar\beta}(0,0)\nn
&&~~~\sim -\pi^2 \left\{G^{-(0)}_{\dot A,-\frac{1}{2}}\left[\sigma^{+\bar\alpha}(w,\bar w)G^{+(0)}_{\dot B,-\frac{1}{2}}\sigma^{-\bar\beta}(0,0)\right]-\sigma^{+\bar\alpha}(w,\bar w)G^{-(0)}_{\dot A,-\frac{1}{2}}G^{+(0)}_{\dot B,-\frac{1}{2}}\sigma^{-\bar\beta}(0,0)\right\}\nn
\eea
Because $G^{-(0)}_{\dot A,-\frac{1}{2}}\sigma^{-\bar\alpha}=0$ and $G^{+(0)}_{\dot A,-\frac{1}{2}}\sigma^{+\bar\alpha}=0$, we get
\bea
&&\bar D^{\bar\alpha}_{\dot A}(w,\bar w)\bar D^{\bar\beta}_{\dot B}(0,0)\nn
&&~~\sim -\pi^2 \left\{G^{-(0)}_{\dot A,-\frac{1}{2}}G^{+(0)}_{\dot B,-\frac{1}{2}}\left[\sigma^{+\bar\alpha}(w,\bar w)\sigma^{-\bar\beta}(0,0)\right]-\sigma^{+\bar\alpha}(w,\bar w)\{G^{-(0)}_{\dot A,-\frac{1}{2}},G^{+(0)}_{\dot B,-\frac{1}{2}}\}\sigma^{-\bar\beta}(0,0)\right\}\nn
\eea
Using the commutation relation
\be
\{ G^{-(0)}_{\dot A,-\frac{1}{2}}, G^{+(0)}_{\dot B,-\frac{1}{2}}\}=\epsilon_{\dot A\dot B}\p
\ee
and the OPE
\be
\sigma^{+\bar\alpha}(w,\bar w)\sigma^{-\bar\beta}(0,0)\sim \epsilon^{\bar\alpha \bar\beta}\frac{1}{|w|^2}
\ee
we get 
\be
\bar D^{\bar\alpha}_{\dot A}(w,\bar w)\bar D^{\bar\beta}_{\dot B}(0,0)\sim-\pi^2\left\{G^{-(0)}_{\dot A,-\frac{1}{2}}G^{+(0)}_{\dot B,-\frac{1}{2}}\epsilon^{\bar\alpha \bar\beta}\frac{1}{|w|^2}-\sigma^{+\bar\alpha}(w,\bar w)\epsilon_{\dot A\dot B}\p\sigma^{-\bar\beta}(0,0)\right\}
\ee
The first term is zero since $G^{+(0)}_{\dot B,-\frac{1}{2}}|0\rangle=0$.
Then we get 
\be
\sigma^{+\bar\alpha}(w,\bar w)\epsilon_{\dot A\dot B}\p\sigma^{-\bar\beta}(0,0)
=-\epsilon_{\dot A\dot B}\epsilon^{\bar\alpha \bar\beta}\p\frac{1}{|w|^2}
\ee
Since
\be  
\p\frac{1}{|w|^2}=-\frac{1}{w^2\bar w}
\ee
we find
\be\label{GNOPE}
\bar D^{\bar\alpha}_{\dot A}(w,\bar w)\bar D^{\bar\beta}_{\dot B}(0,0)
\sim \epsilon_{\dot A\dot B}\epsilon^{\bar\alpha \bar \beta}\frac{\pi^2}{w^2\bar w}
\ee

\section{Perturbations of chiral algebra generators}

Consider a CFT with action
\be
S_0={1\over 4\pi}\int d^2x \, \p_i X^\mu \p^j X^\nu {\cal G}_{\mu\nu}
\ee
 The stress tensor of this unperturbed theory is
 \be
 T^{(0)}_{zz}=- \p_z X^\mu \p_z X^\mu {\cal G}_{\mu\nu}
 \label{fnine}
 \ee
 Consider a small deformation of the theory
 \be
S_0\r S_0+\lambda S_1 ={1\over 4\pi}\int d^2x\,  \p_i X^\mu \p^j X^\nu ({\cal G}_{\mu\nu}+\lambda \t {\cal G}_{\mu\nu})
\label{fone}
\ee
This deformation is described by the $(1,1)$ operator
\be
D={1\over \pi}\p_z X^\mu \p_{\bar z} X^\nu \t  {\cal G}_{\mu\nu}
\ee
The change (\ref{fone})  to the action  gives a change in the stress tensor $T^{(0)}\r T^{(0)}+\lambda T^{(1)}$ with
 \be
 T^{(1)}_{zz}=-  \p_z X^\mu \p_z X^\nu \t  {\cal G}_{\mu\nu}
 \label{ften}
 \ee
 
 Let us work on the plane. Consider first the unperturbed theory. Let there be  a state $|O^{(0)}\rangle$ at $z=0$. We wish to apply the Virasoro operator $L^{(0)}_n$ to this state. We have
 \be
 L^{(0)}_n|O^{(0)}(0)\rangle=\oint_C \frac{dz}{2\pi i} z^{n+1} T^{(0)}_{zz}(z) |O^{(0)}(0)\rangle
 \ee
 The contour $C$ is any contour that circles the origin in a counterclockwise direction, and does not include any other operator insertions.  
 
 Now consider the perturbed theory. At first one might think that the correction to $L_n$ is given by
 \be
 L^{(1)}_n|O^{(0)}(0)\rangle = \oint_C \frac{dz}{2\pi i} z^{n+1}\, T^{(1)}_{zz}(z) |O^{(0)}(0)\rangle ~~(??)
 \label{ftwo}
 \ee
 But the prescription given in \cite{sen} (and the analogous prescription in \cite{gn})  is to take a small circular contour $\hat C$ around the origin, described by
 \be
\hat C: ~~~ |z|=\epsilon
 \ee
 and computing
  \be
 L^{(1)}_n|O^{(0)}(0)\rangle=-\frac{i}{2}\oint_{\hat C} d\bar z z^{n+1} D(z, \bar z) |O^{(0)}(0)\rangle
 \label{fthree}
 \ee
  
 Note that (\ref{ftwo}) involves an integral of a $(2,0)$ operator, while (\ref{fthree}) involves the integral of a $(1,1)$ operator. We would like to understand, in some detail,  the reason why we get (\ref{fthree}) instead of (\ref{ftwo}). The perturbation to the supercharge $\bar G_0$ that has been important for us in this paper is computed by an expression analogous to (\ref{fthree}) so this understanding is relevant. As we will see below, the difference between (\ref{ftwo}) and (\ref{fthree}) arises from a contact term, which is why we have been careful with handling singularities in our  integrals. 

\subsection{The set up}

As noted above, we work in the plane $z$, and place a state $|O^{(0)}\rangle$ at the origin. We place the stress tensor $T(z')$ at the point $z'$; we assume that $|z'| > \epsilon$. We place the state $\langle O_f^{(0)}|$ at infinity. (For simplicity we have not allowed any other operator insertions in the plane.) Thus we have the amplitude
  \bea
A^{(0)}&=&\langle O_f^{(0)}(\infty)|T^{(0)}(z')|O^{(0)}_i (0)\rangle\nn
&=& \langle O_f^{(0)}(\infty)|\int D[X] e^{-S_0} ~T^{(0)}(z')|O_i^{(0)}(0)\rangle
\label{ffive}
\eea
Now we perturb the action as in (\ref{fone}). We do not change the states at $z=0, \infty$ when we perturb the theory; this is indicated by the superscript $(0)$ on these states. The perturbed amplitude is
\be
A^{(0)}+\lambda A^{(1)}
= \langle O_f^{(0)}(\infty)|\int D[X] e^{-S_0-\lambda \int_R d^2 z D} \left ( T^{(0)}(z')+\lambda T^{(1)}(z')\right ) |O_i^{(0)}(0)\rangle
\label{ffiveq}
\ee
The perturbation $D$ is integrated over the whole plane, with the exception of small discs around the initial and final states. These exclusion regions are required to prevent a singularity from the collision of $D$ with the operators defining the states. We can regard these exclusion regions as part of the definition of the states $|O_i^{(0)}(0)\rangle, \langle O_f^{(0)}(\infty)|$. Thus we define the region $R$ as
\be
R: ~~~\epsilon<|z|<{1\over \epsilon}
\ee

The term in (\ref{ffiveq}) at order $\lambda^0$ gives, of course, the quantity $A^{(0)}$. We are interested in the $O(\lambda)$ terms
\bea
A^{(1)}(z')&=&-\langle O_f^{(0)}(\infty)|\int D[X] e^{-S_0} \left (  \int_R d^2 z D(z, \bar z)\right ) \left ( T^{(0)}(z')\right ) |O_i^{(0)}(0)\rangle\nn
&&+\langle O_f^{(0)}(\infty)|\int D[X] e^{-S_0} \left (  T^{(1)}(z')\right ) |O_i^{(0)}(0)\rangle\nn
&\equiv& A^{(1)}_1+A^{(1)}_2
\label{fel}
\eea 
Consider the integral over $z$ in the term $A^{(1)}_1$. We have avoided collisions with the initial and final states by excluding small discs around $z=0, \infty$; this can be considered a part of the definition of these states. But we still have a collision of $D$ with $T^{(0)}$ at $z=z'$. One might define the integration region $R$ to exclude a small disc around $z'$ as well; in fact this is what was done in \cite{sen}. But there is a difficulty with doing this. We do not have any freedom in defining the operator $T$; this operator must be whatever generates diffeomorphisms in the theory, and in our example $T$ is given by (\ref{fnine}),(\ref{ften}). Thus there is no reason why the operator $D(z, \bar z)$ will not collide with the operator $T^{(0)}(z')$. As we will now argue, this collision generates a contact term, which cancels the contribution $A^{(1)}_2$ in (\ref{fel}). This will then illustrate the validity of the prescription used in \cite{sen}  where a hole was cut around $T^{(0)}(z')$ to avoid the collision with $D$, and also the term $A^{(1)}_2$ arising from $T^{(1)}$ was ignored.

\subsection{Regulating the correlators}

To see this cancellation in detail, we must regulate all our correlators. For simplicity we work with just one scalar field
\bea
 S_0+\lambda S_1 &=&{1\over \pi}\int d^2z \p_z X  \p_{\bar z} X ({\cal G}+\lambda \t {\cal G})\nn
 &=&{1\over \pi}\int d^2z \p_z X  \p_{\bar z} X + \lambda {1\over \pi}\int d^2z \p_z X  \p_{\bar z} X 
\label{foneqq}
\eea
The stress tensor is given by
\be
T^{(0)}=-\p_z X \p_z X, ~~~T^{(1)}=-\p_z X \p_z X
\ee

The unperturbed 2-point function of the fundamental field $X$  is 
\be
\langle X(z) X(z')\rangle = -\h \log |z-z'|^2
\label{fthir}
\ee
There is a singularity at $z=z'$, which we regulate as
\be
\langle X(z) X(z')\rangle = -\h \log  |z-z'|^2 \,   h(|z-z'|^2)
\label{fninet}
\ee
where $h(|z|^2)$ goes to zero at $|z|\r 0$ and $h(|z|^2)\r 1$ for $|z|>\delta$ for some small $\delta$. Once we have such a regulator, all the  expressions become well defined even when two operators are coincident. We will use formal expressions that do not mention the regulator, with the assumption that all expressions are really to be understood as those in the presence of the regulator.

From (\ref{fthir}) we have 
\be
\langle \p_z X(z) X(z')\rangle = -\h {1\over z-z'}
\label{fthirq}
\ee
\be
\langle \p_z X(z) \p_z X(z')\rangle = -\h {1\over (z-z')^2}
\label{fthirqq}
\ee
\be
\langle \p_z X (z) \p_{\bar z}X(\bar z')\rangle = -\h \p_{\bar z'}{1\over z-z'}=\h \pi\delta^2(z-z')
\label{ffourt}
\ee
Eq.(\ref{ffourt}) will be the source of the contact term between $T^{(0)}$ and $D$. We regulate  these operators as
\be
T^{(0)}(z')=-\p_z X(z'+\epsilon)\p_z X(z')-{1\over 2\epsilon^2}, ~~~~D(z, \bar z)=\frac{1}{\pi}\p_zX(z+\epsilon) \p_{\bar z}X(\bar z)
\ee

\subsection{The contact term}

We can now compute the contact terms arising in the integration
$
\int_R d^2 z D(z, \bar z) T^{(0)}(z')
$.
The correlator is given by a sum of  Wick contractions. The factor $\p_{\bar z} X$ in $D$ can contract with one of the two $\p_zX$ factors in $T^{(0)}$; the two possibilities give a factor $2$.  The contraction gives a factor $\h \pi\delta^2(z-z')$. The factors left over are $\p_z X\p_z X$, which are equivalent to  $T^{(1)}(z')$. Thus the contact term is
\be
\int_R d^2 z D(z, \bar z) T^{(0)}(z')\r -\int_R d^2 z \delta^2(z-z') \p_z X  \p_z X = - \p_z X(z')  \p_z X(z')=T^{(1)}(z')
\label{fsfive}
\ee
Using this in (\ref{fel}) we see that the contact term in $A^{(1)}_1$ cancels the term $A^{(1)}_2$.\footnote{There is an ambiguity in the normal ordering constant produced in this process, so the cancellation can leave over a term proportional to the identity operator.  We will define Virosoro modes $L^{(1)}_n$ below; in these modes this identity operator can shift the zero mode by a constant $L^{(1)}_0\r L_0^{(1)}+const$. This is a standard ambiguity in the definition of Virasoro generators, which leads to an arbitrary parameter $\alpha$ in the anomaly term  ${c\over 12} (m^3-\alpha m)$. By convention, we shift the constant to get $\alpha=1$.}

\subsection{The remaining terms}

Having dealt with the contact term, we can write the correlator of $T^{(0)}(z')$ with the other operators in the plane using the usual Ward identity. We have
\bea
-\langle O_f^{(0)}(\infty)| \int_R d^2 z D(z, \bar z) T^{(0)}(z')|O_i^{(0)}(0)\rangle&=&\nn
&&\hskip-100pt -\langle O_f^{(0)}(\infty)| \int_R d^2 z \left ({D(z, \bar z)\over (z'-z)^2}+{\p_z D(z, \bar z)\over z'-z}\right ) |O_i^{(0)}(0)\rangle\nn
&&\hskip-100pt -\langle O_f^{(0)}(\infty)| \int_R d^2 z D(z, \bar z) \left ( \sum_n (z')^{-n-2} L^{(0)}_n|O_i^{(0)}(0)\rangle \right )
\nn
 &&\equiv Q_1+Q_2
 \label{ftone}
\eea

\subsection{The term $Q_1$}

For the term $Q_1$  in (\ref{ftone}) we get
\bea
Q_1&=&-\langle O_f^{(0)}(\infty)| \int_R d^2 z \left ({D(z, \bar z)\over (z'-z)^2}+{\p_z D(z, \bar z)\over z'-z}\right ) |O_i^{(0)}(0)\rangle\nn
&=& -\langle O_f^{(0)}(\infty)| \int_R d^2 z \p_z \left ({D(z, \bar z)\over z'-z}\right ) |O_i^{(0)}(0)\rangle\nn
&=& -\langle O_f^{(0)}(\infty)| \frac{i}{2}\int_{\p R} d\bar z   \left ({D(z, \bar z)\over z'-z}\right ) |O_i^{(0)}(0)\rangle
\eea
Let us now analyze the contributions from the various components of the boundary $\p R$:

\b

(a) The boundary $\p R$ has a circle at $|z|={1\over \epsilon}$. The contribution here described the Virasoro generators acting on the state at $z=\infty$. We will focus on the Virasoro generators acting on the state $|O^{(0)}\rangle$ at $z=0$, so we will ignore this contribution in what follows.

\b

(b) There is a boundary at $|z|=\epsilon$. The contribution here is 
\be
 -\langle O_f^{(0)}(\infty)| \frac{i}{2}\oint_{|z|=\epsilon} d\bar z   \left ({D(z, \bar z)\over z'-z}\right ) |O_i^{(0)}(0)\rangle
 \label{fsixt}
 \ee
 This is the part that will give the correction $L^{(1)}_n$ on the state at $z=0$.

 \b
 
 (c) The question now is whether we have an additional component of $\p R$ at $|z-z'|=\epsilon$. In \cite{sen} a hole of radius $\epsilon$ was cut around $z'$ to avoid the contact term, so we indeed had a boundary at $|z-z'|=\epsilon$. But we have argued that we do not a priori have the freedom to cut out such a hole, and we have kept all contact terms that arise from integrating $z$ across the location $z'$. We will now see that there is in fact no additional boundary contribution at  $|z-z'|=\epsilon$. To see this, note that the terms in the $T^{(0)}(z') D(z, \bar z)$ OPE arise from contracting a factor $\p_z X$ in $T^{(0)}$ with  a factor $\p_z X$ in $D$. We can write 
 \be
 D(z, \bar z)=\p_z X(z+\epsilon) \p_{\bar z} X(z) = \p_z \Big ( X(z+\epsilon) \p_{\bar z} X(z)\Big ) - X(z+\epsilon)\, \p_z\p_{\bar z} X(z)
 \label{ftwenty}
 \ee
 We now regard all correlators as everywhere finite quantities defined through the regulator (\ref{fninet}). Then the first term in (\ref{ftwenty}) generates an integrand which is a total derivative $\p_z$ of  an everywhere finite integrand. Thus the integral $\int d^2 z$ of the corresponding contribution  gives boundary terms only at  $|z|={1\over \epsilon}$ and $|z|=\epsilon$. In the second term in (\ref{ftwenty}) the factor $X(z+\epsilon)\, \p_z\p_{\bar z} X(z)$ vanishes by the equation of motion. This would not be true at points where there was a contact interaction with a $\p_{\bar z} X$, but the Wick contraction giving this contact term has already been included in the computation (\ref{fsfive}), and so there are no such contact terms to consider anymore.  
 
 \b
 
 Let us collect all the terms we found above for the amplitude $A^{(1)}$ defined in (\ref{fel}). As noted, we will focus on the terms giving the Virasoro generators on the state at $z=0$; there are similar terms for the generators acting at $z=\infty$ that we do not list. We denote by $\t A^{(1)}$ the part of $A^{(1)}$ that is relevant for the action on the state at $z=0$. We have     \bea
 \t A^{(1)}(z')&=& -\langle O_f^{(0)}(\infty)|\frac{i}{2} \oint_{|z|=\epsilon} d\bar z   \left ({D(z, \bar z)\over z'-z}\right ) |O_i^{(0)}(0)\rangle\nn
 &&-\langle O_f^{(0)}(\infty)| \int_R d^2 z D(z, \bar z) \left ( \sum_n (z')^{-n-2} L^{(0)}_n|O_i^{(0)}(0)\rangle \right )
 \label{ftthree}
  \eea
  where the second term in the quantity $Q_2$ from (\ref{ftone}). 
  
  \subsection{Defining the modes $L_n$}
  
 The quantity $A^{(1)}$ defined in (\ref{fel}) has the stress tensor inserted at a particular point $z'$. We can integrate over $z'$ to define modes $L_n$. Thus we define
 \be
 \t A^{(1)}_n=\oint \frac{dz'}{2\pi i} z'^{n+1}\t A^{(1)}(z')
 \ee
   where the $z'$ contour circles the origin counterclockwise, and lies at $|z'|>\epsilon$. In the first term on the RHS of (\ref{ftthree}) we expand as $(z'-z)^{-1}=\sum_{m=0}^\infty  z^m z'^{-1-m}$. We get
 \bea
  \t A^{(1)}_n&=& -\langle O_f^{(0)}(\infty)| \frac{i}{2}\oint_{|z|=\epsilon} d\bar z  z^{n+1} D(z, \bar z) |O_i^{(0)}(0)\rangle\nn
  &&-~ \langle O_f^{(0)}(\infty)| \int_R d^2 z D(z, \bar z)  L^{(0)}_n|O_i^{(0)}(0)\rangle \nn
  \label{ftfive}
  \eea
  The second term on the RHS (\ref{ftfive}) can be identified as the contribution where $L^{(0)}_n$ acts on $|O_i^{(0)}(0)\rangle$ and the resulting state is evolved by using the first order correction to the Hamiltonian. The first term on the RHS of (\ref{ftfive}) must then be identified as the action of $L^{(1)}_n$ on $|O_i^{(0)}(0)\rangle$; the resulting state is then evolved by the leading order Hamiltonian. Thus we define
  \be\label{App L1}
 L^{(1)}_n |O_i^{(0)}(0)\rangle =-\frac{i}{2} \oint_{|z|=\epsilon} d\bar z  z^{n+1} D(z, \bar z) |O_i^{(0)}(0)\rangle
  \ee
  which is the prescription in \cite{sen, sen2}. 

\section{Defining $L^{(1)}_n$ for general deformations}\label{appd}
 
 In the above simple example we saw that the contribution to $L^{(1)}$ from the $(2,0)$ tensor $T^{(1)}$ was cancelled by a contact term, and we were left with an integral over the $(1,1)$ operator $D$.
 
 We might now wonder if such a cancellation will  arise for all marginal deformations. The difficulty with addressing this question is that while we are always given a $(1,1)$ operator $D$ defining the deformation, we do not always know the analogue of the $(2,0)$ operator $T^{(1)}$. For example in the orbifold CFT, the deformation operator $D$ is given by (\ref{D 1/4}).  The deformation of the theory is defined by 
 \be
 S_0\r S_0+\lambda S_1=S_0+\lambda \int d^2 z D(z, \bar z)
 \label{ftheight}
 \ee
 We are not, however, given an action of the type $S[X,g]$ which depends on the field theory variables $X$ and also allows an arbitrary metric $g$ on the $z$ space. If we {\it did} have such an action, then we could compute the stress tensor as 
 \be
 T_{ij}= {2\over \sqrt{g}}{\delta S\over \delta g^{ij}}=T^{(0)}_{ij}+\lambda T^{(1)}_{ij}
 \ee
 and we would obtain $T^{(1)}_{zz}$. But we are given a prescription for computing the deformation only once we go to coordinates where the metric is conformal
 \be
 ds^2=e^{\phi(z, \bar z)} dzd\bar z
 \label{fthone}
 \ee
 In such coordinates we integrate over $\int d^2 z D(z, \bar z)$. (This integral is independent of the conformal factor $\phi$ since $D$ is a $(1,1)$ operator.) But since we cannot study the deformation in coordinates which have a metric componet $d\bar z^2$, we cannot find $T^{(1)}_{zz}$.

 \subsection{Defining $L_n$}\label{define Ln}
 
 If we cannot find $T^{(1)}$, how should we proceed to find the $L^{(1)}_n$? Instead of defining the $L_n$ through an integral over $T$, we can define them through their effect, which is to generate a diffeomorphism. First we consider the abstract definition of the $L_n$ in a general theory;\footnote{A detailed discussion of the definition of the $L_n$ is given in \cite{wong}. A Hamiltonian derivation of $L^{(1)}_n$ is given in \cite{ovrut}.} in the following subsection we will consider the infinitesimal deformation of the theory. 
 
 Consider correlators on a surface $\Sigma$. We take $\Sigma$ to have the topology of a sphere, and further use the flat metric $g_0$ on the plane 
 \be
g_0: ~~ ds^2=dz d\bar z
\ee
   We cut holes on this sphere with boundaries given by curves $C_i$. At the edge of each hole we have a state $|O_i\rangle$.  The path integral with action $S_0$ generates the amplitude
 \be
 A=\langle O_1 \dots O_k\rangle_{g_0}
 \label{fthfour}
 \ee

 Now take a region $R'$ that does not overlap with the holes. In this region we make an infinitesimal change of the metric $g_0\r g_0+\delta g$
 \be
ds^2\r dzd\bar z+h(z, \bar z) d\bar z^2
\label{fthsix}
\ee
This gives a new amplitude
\be
A+\delta A = \langle O_1 \dots O_k\rangle_{g_0+\delta g}
\label{fthtwo}
\ee
We now ask: can the amplitude (\ref{fthtwo}) be expressed in a way where the metric is returned back to the metric $g_0$ in (\ref{fthfour}) but the states $|O_i\rangle$ are altered? The fact that we can do this is the Ward identity, and the alteration of the states will define the Virasoro generators. We first remove the $g_{\bar z\bar z}$ part of the metric by a diffeomorphism 
\be
z\r z+\epsilon^z
\label{fthseven}
\ee
This diffeomorphism generates the change
\be
g_{\bar z\bar z}\r g_{\bar z\bar z} -2\epsilon_{\bar z, \bar z} = g_{\bar z\bar z} - \epsilon^z{}_{, \bar z}
\label{fthfive}
\ee
where we have used $g_{z\bar z}=\h, g^{z\bar z}=2$. Setting
\be
 \epsilon^z{}_{, \bar z}=h(z,\bar z)
\ee
in (\ref{fthsix}), we remove the $d\bar z^2$ part of the metric. We do however generate a new part $h_{z\bar z}=-\epsilon_{\bar z, z}=-\frac{1}{2}\epsilon^z{}_{,z}$. Due to invariance under the conformal factor $\phi$ in (\ref{fthone}), we ignore this change, and set the metric to again be $ds^2=dzd\bar z$. 

While the metric has returned to the flat one, the curves $C_i$ have been deformed by the diffeomorphism (\ref{fthseven}) to new curves $C'_i$. Since the support of $h(z, \bar z)$ was confined to a region $R'$ that did not overlap with the curves $C_i$, we have $\epsilon^z_{, \bar z}=0$ in the neighborhood of the $C_i$; i.e., $\epsilon^z$ is holomorphic near the $C_i$. 

We wish to leave the state at the hole unchanged. Since the curve $C_i$ has changed to $C'_i$, how should we define the `same' state on this new curve? First recall how we place a state on the curve $C_i$. We consider an auxiliary space with coordinate $\t z$, and place sources in the region $|\tilde z|<1$ to create a state $|O_i\rangle$ at the circle $|\tilde z|=1$. On the surface $\Sigma$, we take a coordinate $z_i$ around the curve $C_i$ such that $C_i$ is given by $|z_i|=1$. We then glue the state $|O_i\rangle$ to the curve $C_i$ by identifying the points on the circles $|\t z|=1$ and $|z_i|=1$. 

To glue in  the same  state after the curve has been deformed to $C'_i$, we keep the construction in the auxiliary space $\t z$ unchanged. On the surface $\Sigma$ we have the coordinate $z'_i=z_i+\epsilon^z$. We now identify the circle  $|\t z|=1$ with the circle $|z'_i|=1$.

While we have now placed the `same' state $|O_i\rangle$ as before at the $i$th hole, the points on $\Sigma$ that define the hole are now different, as the points on $C'_i$ are different from the points on $C_i$. We can however, evolve the state $|O_i\rangle$ radially outwards or inwards, as needed, till we arrive at a state $|\t O_i\rangle$ on the original curve $C_i$. Since the diffeomorphism is small, we have
\be
|\t O_i\rangle=|O_i\rangle+\epsilon'\delta |O_i\rangle
\label{ffone}
\ee
where $\epsilon'$ is small. The diffeomorphism itself can be expanded as
\be
\epsilon^z=\epsilon'\sum_n a_n z^{n+1}
\label{ffthree}
\ee
Then we write
\be
\delta |O_i\rangle=\sum_n a_n L_n|O_i\rangle
\label{ffFour}
\ee
and this defines the action of the operators $L_n$ acting on the state at the $i$th hole. Note that we do not need an explicit expression for the stress tensor in this definition. 

\subsection{Deforming the theory}

Now consider the deformation of the theory (\ref{ftheight}). The first order correction to the amplitude $A^{(0)}$ is given by
\be\label{A(1) R}
A^{(1)}= -\langle \int_R d^2 z D(z, \bar z) O^{(0)}_1 \dots O^{(0)}_k\rangle_{g_0}
\ee
We have states  $|O^{(0)}_i\rangle$ defined on the curved $C_i$. The superscript $(0)$ denotes, as before, that these are unperturbed states; we will not alter the states when we deform the theory by the addition of $\lambda D$ to the Lagrangian. 

The region $R$ is defined as follows. We have to exclude the holes described by the $C_i$. We also exclude a small strip of proper width $\epsilon$ around these holes; this is to prevent the collision of $D$ with the operators defining the holes. The region outside these strips we take to be $R$, the domain of integration of $D$.

Now we wish to consider the change of metric (\ref{fthsix}). With the metric in this new form, we do not have a prescription to compute the new amplitude. But we apply the diffeomorphism (\ref{fthseven}) to bring the metric back to the form (\ref{fthone}), and then we can again find the amplitude by integrating the position of $D$. 

The diffeomorphism leads to the effect noted in section \ref{define Ln}. 
The curves $C_i$ move to curves $C'_i$. The states $|O^{(0)}_i\rangle$ at the hole are left unchanged, but these are now defined on different curves $C'_i$ on $\Sigma$. We must now evolve the state back to the original curves $C_i$, to get the correction (\ref{ffone}).

This evolution between the two curves $C_i$ and $C'_i$ must, of course, be done using the complete action $S_0+\lambda S_1$. First consider the part we get from using just the leading order action $S_0$. The corresponding effect on the state gives $L^{(0)}_n|O^{(0)}_i\rangle$. The integration $\int_R d^2 z D$ gives the first order correction to the evolution in the region between the curves $C_i$. 


Next, consider the contribution from the change of the integral region  $R$ in eq. (\ref{A(1) R}). Before the coordinate transformation, the region $R$ is bounded by $C_{i}$ and and after  the coordinate transformation the region $R'$ is bounded by $C'_{i}$. 
The extra contribution arising from the  integral  $\int d^2 z D$ in the strip between these curves  is
\be
\oint_{C_{i}} (\epsilon^{\tau}d\sigma-\epsilon^{\sigma}d\tau)D=\oint_{C_{i}} \frac{i}{2}(\epsilon^{z}d\bar z-\epsilon^{\bar z}dz)D
\ee
Since the diffeomorphism has only the component $\epsilon^z$, this becomes
\be
\frac{i}{2}\oint_{C_{i}}\epsilon^{z}d\bar zD
\ee
Thus from eq. (\ref{A(1) R}), we found the contribution to $\delta |O^{(0)}_{i}\rangle$ is
\bea
-\frac{i}{2}\oint_{C_{i}} d\bar z \epsilon^{z} D(z,\bar z)|O^{(0)}_{i}\rangle
\eea
Using $\epsilon^{z}=\epsilon' z^{n+1}$, we get 
\be
L^{(1)}_n|O^{(0)}\rangle=-\frac{i}{2}\oint_C d\bar z z^{n+1} D(z,\bar z)|O^{(0)}\rangle
\label{fffive}
\ee
which is the same as (\ref{App L1}).

\subsection{The choice of $C$}

In the above discussion the states were described using contours $C_i$ on $\Sigma$. This contour shows up in the definition of the Virasoro generators (\ref{fffive}). We have also had to use a regulator to separate the domain of integration of $D$ from the contour where the state was defined. We did this by taking a thin band of proper width $\epsilon$ around the $C_i$ as a region where the integration of $D$ would not expend. But one could have chosen a differently shaped band, and this would change the definition of the Virasoro generators. What is the significance of these possible ambiguities in our definition?

In \cite{sen,sen2} the deformation $D$ was used to study background independence of string field theory: the action of $D$ moved the world sheet theory to a slightly different one corresponding to a slightly different background, and one then computed the correlators of the `same' states $|O^{(0)}_i\rangle$ in this deformed theory. There was however no canonical definition of the `same' state in different theories; while a certain prescription has been used to keep $|O^{(0)}_i\rangle$ unchanged, we find that the regulators needed in carrying out the computations with this state introduce an arbitrariness in how the state changes as we change the theory. This arbitrariness is not a difficulty for the study of background independence, as different choices of how the `same' state is defined as we move the theory correspond to different connections on the space of theories. 

In our computations on the other hand the goal is to find the lift of the dimensions of certain primary operators. We need a well defined result, which should be consistent with the perturbation theory in the path integral formalism.
 Thus let us consider the ambiguities appearing in the definition (\ref{fffive}). Let the operator $O$ have dimensions $(h, \bar h)$ and be inserted at a point $z=0$. We let the contour be a small curve surrounding $z=0$, so that we can expand the integrand in (\ref{fffive}) using the OPE
\be
D(z, \bar z) O^{(0)}(0)\sim \sum_{m,n} \t O_{m,n} z^{m-1} \bar z^{n-1}
\label{fsone}
\ee
where $\t O_{m,n}$ has dimensions 
\be
(\t h, \t {\bar h})=(h+m, \bar h+n)
\ee
For the integral in the $L_{0}^{(1)}$, we have
\be
\oint_{C_{0}} \frac{d\bar z}{\bar z}z^m\bar z^{n}
\ee
We now consider several cases:

\b

(i)  Suppose $m+n\ge 1$. Then as we shrink the size of the contour we find
\be
\oint_{C_{0}} \frac{d\bar z}{\bar z}z^m\bar z^{n}\r 0
\ee
independent of the shape of the contour. Thus operators $\tilde O_{m,n}$ of higher energy act as irrelevant operators for the computation of the lift.

\b

(ii)  Suppose $m+n= 0$. Then as we shrink the size of the contour
\be
\oint_{C_{0}} \frac{d\bar z}{\bar z}z^m\bar z^{n}\r {\rm finite}
\ee
but the value of this finite quantity depends on the shape of the curve $C_0$ . We get a shape independent result in the special case
 $\int {d\bar z\over \bar z}=2\pi i$. But consider the integral $\int{d\bar z\over z}$. Suppose the contour  is a rectangle with corners $(\tau,\sigma)=(\pm\epsilon, \pm\epsilon')$. Then 
 \be
 \oint_{C_{0}} {d\bar z \over z}=2 \log { \epsilon+i\epsilon'\over \epsilon-i\epsilon'}-2\log { \epsilon'+i\epsilon\over \epsilon'-i\epsilon}
 \ee
 and this depends on the ratio ${\epsilon\over \epsilon'}$. 

\b

(iii) Suppose $m+n\le -1$. Then the integral diverges as $\epsilon^{m+n}$ in general as we shrink the curve $C_{0}$. Different shapes of the curves lead to different value of the integral. 

\b

As we can see from the above, the last two cases lead to some ambiguities in the definition of the $L^{(1)}_{n}$, and in particular for $L^{(1)}_0$. The computation of $L^{(1)}_0$ is similar to that for the operator $\bar G ^{(1)}_0$ that arises in the lifting of energies that we have been interested in. So let us analyze these ambiguities in more detail.

We are interested in states that are Ramond ground states on the right; i.e., $\bar h={c\over 24}$, the lowest possible value for the right sector. Thus in the expanion (\ref{fsone}) we have
$n\ge 0$. But the left sector of our states is excited to an arbitrarily high level, so $m$ can be arbitrarily negative, and then we see that in cases (ii) and (iii) above there is an ambiguity arising from the exact shape of the curve.

We will now observe that the correct way to remove this ambiguity is to take the curve to be an exact circle in the $z$ plane; the corresponding curve on the cylinder is a circle at constant $\sigma$. To see this, first recall that in deriving the lifting from the path integral in \cite{hmz} we used that fact that the only states propagating between $\tau=\pm{T\over 2}$ had energy $E\ge E^{(0)}$. We had taken the region of integration for the deformation operators to extend from $\tau=-{T\over 2}+\epsilon$ to $\tau={T\over 2}+\epsilon$; i.e., the curves bounding the domain of integration were exact circles. But now suppose we had taken curves $\t C$ of some other shape. Then the integration of one or both of the deformation operators in the region  $\tau\approx -{T\over 2}$ and $\tau\approx {T\over 2}$ can produce states where the right dimension is still $\bar h'={c\over 24}$ but the left dimension is $h'<h$. These states have $E<E^{(0)}$, and their contribution to the amplitude grows faster than the contribution $Te^{-E^{(0)}T}$ that we use to extract $\delta^{(2)} \bar h$. Thus the expression we used for the lift would not be correct.

But suppose we did take the curves bounding the domain of integration to be circles at constant $\tau$. Then we can do the integral over $\sigma$ first, which sets the momentum of the propagating state to equal the momentum $h-\bar h=h-{c\over 24}$ of the initial state $|O^{(0)}\rangle$. Since $\bar h'\ge {c\over 24}$, we see that $h'\ge h$, and thus the energy of the propagating state is $E\ge E^{(0)}$. On the plane, the corresponding simplification is expressed by
\be
\oint_{C_{0}} \frac{d\bar z}{\bar z}z^m \bar z^n=2\pi  i\epsilon^{2m} \delta_{m,n}
\ee
We see that since $m\ge 0$, there is no singular term in our computations when the curve is taken as a circle.

The expression for the lift that we use is then applicable. Thus we see that the choice of curve over which $D$ or $\bar D$ is integrated is not arbitrary, but dictated by the fact that we are using the resulting construction to compute the lift (and not just to find an arbitrary connection on the space of conformal theories).

\end{document}